\documentclass[11pt,a4paper]{amsart}
\usepackage[margin=2cm]{geometry}
\usepackage[title]{appendix}
\usepackage{float}

\usepackage[backend=biber, sorting=nyt, style=numeric]{biblatex}
\usepackage{csquotes}

\usepackage[normalem]{ulem}

\usepackage[foot]{amsaddr}
\makeatletter
\renewcommand{\email}[2][]{%
  \ifx\emails\@empty\relax\else{\g@addto@macro\emails{,\space}}\fi%
  \@ifnotempty{#1}{\g@addto@macro\emails{\textrm{(#1)}\space}}%
  \g@addto@macro\emails{#2}%
}
\makeatother

\usepackage{amsmath}
\usepackage{amsfonts}
\usepackage[english]{babel}
\usepackage{graphicx} 
\usepackage{float}
\usepackage{tikz}
\usetikzlibrary{arrows, shapes, backgrounds}
\usepackage{xcolor}
\usepackage{framed}
\usepackage{enumitem}
\usepackage{adjustbox}

\usepackage{pgfplotstable}
\usepackage{booktabs}

\usepackage{array}
\usepackage{colortbl}
\usetikzlibrary{er, positioning}

\bibliography{sec-mat-models.bib}

\title{Pricing cyber-insurance for systems via maturity models} 

\author{Henry R.K. Skeoch$^{\ast1}$}
\address{$^1$University College London, Gower Street, London, United Kingdom}

\email{henry.skeoch.19@ucl.ac.uk}

\author{David Pym$^{1,2}$}
\address{$^2$Institute of Philosophy, University of London}
\email{d.pym@ucl.ac.uk}

\address{$^{\ast}$Corresponding author}

\pgfplotsset{compat=1.17}

\begin{document}
\newcommand{\payrollerd}{
    \node[] (Operations) at (-5,-2) {Operations (1)};
    \node[] (Service) at (-5, -6) {Service (2)};
    \node[] (Systems) at (-5, -10) {Systems (3)};
    \draw[dashed] (-6,-4.875) -- (11,-4.875);
    \draw[dashed] (-6,-7.5) -- (11,-7.5);
    \draw[dashed] (-3.5, 1.5) -- (-3.5, -12);
    \node[entity] (finance) at (0,-2) {Finance};
    \node[attribute] (department-finance) at (0,-1) {Department}
        edge(finance);
    \node[entity, text width=2cm, align=center] (hr) at (8,-2) {Human Resources};
     \node[attribute] (department-hr) at (8,-1) {Department}
        edge(hr);
    \node[entity] (employees) at (4,-2) {Employees};
     \node[attribute, double] (Resource) at (4,-1) {Resource}
        edge(employees);
    \node[entity] (records) at (4,-6) {Records};
    \draw[] (3.25, -6) -- (4, -6.375);
    \draw[] (4,-6.375) -- (4.75,-6);
    \draw[] (4.75, -6) -- (4, -5.625);
    \draw[] (4, -5.625) -- (3.25, -6);
     
    \node[attribute] (data) at (5.5,-7) {\underline{Data}}
        edge(records);
    \node[entity, dashed] (payroll) at (4, -9.75) {Payroll};
    \node[attribute] (database) at (4, -10.5) {Database}
        edge(payroll);
    
    \node[relationship] (maintains) at (8, -6) {maintains}
        edge(hr)
        edge(records);
        
    \node[relationship] (pays) at (2,-2) {pays}
        edge(finance)
        edge(employees);
        
    \node[relationship] (storedin) at (4, -8.25) {stored in}
        edge(records)
        edge(payroll);
    
    \node[relationship] (represent) at (4, -3.75) {represent}
        edge(employees)
        edge(records);
        
    \node[relationship] (reads) at (0, -6) {reads}
        edge(finance)
        edge(records);
}

\newcommand{\instantiateerd}{
    \tikzset{every entity/.style={text width=2.5cm, align=center, fill=white, scale=0.8}}
     \tikzset{every relationship/.style={text width=1.5cm, align=center, fill=white, scale=0.6}}
      \tikzset{every attribute/.style={align=center, fill=white, scale=0.7}}
    \node[] (Operations) at (-4,-2) {Ops. (1)};
    \node[] (Service) at (-4, -6) {Serv. (2)};
    \node[] (Systems) at (-4, -10) {Sys. (3)};
    \draw[dashed] (-5,-3) -- (10,-3);
    \draw[dashed] (-5,-7.5) -- (10,-7.5);
    \draw[dashed] (-3, 1.5) -- (-3, -12);
}

\newcommand{\erdshading}{
    \begin{scope} [on background layer]
    \node[] (Criticality) at (1.75, -11) {Criticality};
    \node[] (Sensitivity) at (6.25, -11) {Sensitivity};
    \begin{scope} [fill opacity = 0.2]
    \fill[red] (2,-5) ellipse (4 and 6.5);
    \fill[blue] (6, -5) ellipse (4 and 6.5);
    
    \end{scope}
    \end{scope}
}

\newcommand{\onlineretailerd}{
    \node[entity] (customerserviceops) at (0.25,-1.5) {Customer Service Operations};
    
    \node[entity] (delivery) at (-0.5,-5.5) {Delivery};

    \node[entity] (presentation-server) at (0.25,-8.75) {Presentation Server};
    \node[attribute] (website) at (0.25, -10) {Website}
        edge(presentation-server);

    \node[entity] (stock-control) at (4,-2) {Stock Control};
    \node[entity] (inventory) at (4,-5.5) {Inventory};
    \node[attribute] (warehouse) at (5.25, -6.25) {Warehouse}
        edge(inventory);
    
    \node[entity] (stock-database) at (4,-8.5) {Stock Database};

    \node[entity] (customer-interaction) at (7.75,-1.5) {Customer Interaction};
    \node[entity, text width=3cm] (customer-comms) at (8.25,-5.5) {Communications};
    \node[attribute] (SMS) at (8, -6.5) {SMS}  
        edge(customer-comms);
    \node[attribute] (Email) at (6.5, -7) {Email}
        edge(customer-comms);
    
    \node[entity] (customer-data) at (7.75,-8.5) {Customer Database};

    \node[attribute] (department) at (4, -1) {Department}
        edge(customerserviceops)
        edge(stock-control)
        edge(customer-interaction);
    \node[relationship] (read-write1) at (6, -9.75) {read and write}
        edge(presentation-server)
        edge(customer-data);

    \node[relationship] (read-write2) at (2.25, -7.25) {read and write}
        edge(presentation-server)
        edge(stock-database);

    \node[relationship] (records) at (4, -6.875) {records}
        edge(inventory)
        edge(stock-database);

    \node[relationship] (manage1) at (4, -3.75) {manages}
        edge(stock-control)
        edge(inventory);

    \node[relationship] (handles) at (9,-4) {handles}
        edge(customer-interaction)
        edge(customer-comms);

    \node[relationship] (oversee) at (-1, -4) {oversees}
        edge(customerserviceops)
        edge(delivery);

    \node[relationship] (exchange-instruction) at (0, -6.875) {exchange data}
        edge(delivery)
        edge(presentation-server);

    \node[relationship] (use) at (9, -6.875) {use}
        edge(customer-comms)
        edge(customer-data);

    \node[relationship] (correspond) at (4, -0) {liase}
        edge(customerserviceops)
        edge(customer-interaction);

    \node[relationship] (transfer) at (1.5, -4) {transfer}
        edge(delivery)
        edge(inventory);

    \node[relationship] (informs) at (7, -4) {informs}
        edge(delivery)
        edge(customer-comms);
    \erdshading

    }

    \newcommand{\bankingerd}{
     \node[entity] (payments)  at (0.25,-1.25) {Payments};
    \node[entity] (payment-infrastructure)  at (-0.5,-4.5) {Payment Infrastructure};
    \node[entity] (ledger) at (1,-9.5) {Ledger};

    \node[entity] (bookkeeping) at (4,-1.5) {Bookkeeping};
    \node[entity] (back-end) at (4,-5.375) {Back-end interface};
    \node[entity] (backend-db) at (4,-8.5) {Accounts Database};

    \node[entity] (customer-interaction) at (7.75,-1.25) {Customer Interaction};
    \node[entity, text width=3cm] (customer-interfaces) at (8.5,-4.5) {Customer Interfaces};
    \node[attribute] (app) at (9.25, -6.25) {App}
        edge(customer-interfaces);
    \node[attribute] (website) at (8.75, -7) {Website}
        edge(customer-interfaces);    
    \node[entity] (customer-data) at (7.75,-8.5) {Customer Data};

    \node[attribute] (department) at (4, 0) {Department}
        edge(payments)
        edge(bookkeeping)
        edge(customer-interaction);

    \node[relationship] (read-modify) at (6.675, -6.5) {read and modify}
        edge(customer-interfaces)
        edge(customer-data);
    \node[relationship] (read1) at (6, -10) {read}
        edge(customer-data)
        edge(backend-db);

    \node[relationship] (updates) at (2.25, -7.5) {updates}
        edge(ledger)
        edge(back-end);

    \node[relationship] (writes-to1) at (-0.5, -7.5) {writes to}
        edge(payment-infrastructure)
        edge(ledger);

    \node[relationship] (writes-to2) at (4, -6.75) {writes to}
        edge(back-end)
        edge(backend-db);

    \node[relationship] (manages1) at (0, -3) {manages}
        edge(payments)
        edge(payment-infrastructure);

    \node[relationship] (manages2) at (4, -3) {manages}
        edge(bookkeeping)
        edge(back-end);

    \node[relationship] (manages3) at (8, -3) {manages}
        edge(customer-interaction)
        edge(customer-interfaces);

    \node[relationship] (interact1) at (2.125, -4.25) {send and receive}
        edge(customer-interfaces)
        edge(payment-infrastructure);
        }

\newcommand{\pharmaerd}{
\node[entity] (logistics) at (0.25,-2) {Logistics};
    \node[entity] (distribution) at (0.25,-5) {Distribution};
    \node[entity] (logistics-db) at (0.25,-8.5) {Logistics Database};

    \node[entity] (factory) at (4,-2) {Factory};
    \node[entity] (manuf-process) at (4,-5) {Manufacturing Processes};
    \node[entity] (manuf-system) at (4,-8.5) {Machinery};

    \node[entity] (randd) at (7.75,-2) {Research \& Development};
    \node[entity, text width=3cm] (communications) at (7.75,-5) {IT services};
    \node[entity] (sensitive-data) at (7.75,-8.5) {Sensitive Data};

    \node[attribute] (email) at (9, -6) {Email}
        edge(communications);
    \node[attribute] (filesystems) at (8.5, -7) {Filesystems}
        edge(communications);
    
    \node[attribute] (departments) at (4, 0) {Departments}
        edge(logistics)
        edge(factory)
        edge(randd);


    \node[relationship] (use) at (4, -3.25) {use}
        edge(communications)
        edge(logistics)
        edge(factory)
        edge(randd);

    \node[relationship] (access-update) at (6.25, -6.75) {access and update}
        edge(communications)
        edge(sensitive-data);

    \node[relationship] (read) at (4.75, -6.75) {read}
        edge(manuf-process)
        edge(sensitive-data);

    \node[relationship] (instruct) at (3.5, -6.75) {instruct}
        edge(manuf-process)
        edge(manuf-system);

    \node[relationship] (controls) at (2.75, -3.75) {controls}
        edge(factory)
        edge(manuf-process);

    \node[relationship] (manages) at (0, -3.75) {manages}
        edge(logistics)
        edge(distribution);

    \node[relationship] (uses-and-maintains) at (-0.5, -6.5) {uses and maintains}
        edge(distribution)
        edge(logistics-db);

    \node[relationship] (receives output from) at (1.5, -6.5) {receives output from}
        edge(distribution)
        edge(manuf-system);

    \node[relationship] (makes-products) at (6.5, 0) {makes drugs designed by}
        edge(randd)
        edge(factory);

    \node[relationship, text width=3cm] (coordinate) at (1.25,0) {coordinate}
        edge(logistics)
        edge(factory);
}

\maketitle

\begin{abstract}
Pricing insurance for risks associated with information technology systems presents a complex modelling challenge, combining the disciplines of operations management, security, and economics. This work proposes a socioeconomic modelling framework for cyber-insurance decisions compromised of entity relationship diagrams, security maturity models, and economic models, addressing a long-standing research challenge of capturing organizational structure in the design and pricing of cyber-insurance policies. Insurance pricing is usually informed by the long experience insurance companies have of the magnitude and frequency of losses that arise in organizations based on their size, industry sector, and location. Consequently, their calculations of premia will start from a baseline determined by these considerations. A unique challenge of cyber-insurance is that data history is limited and not necessarily informative of future loss risk meaning that established actuarial methodology for other lines of insurance may not be the optimal pricing strategy. The modelling framework proposed in this paper provides a vehicle for agreement between practitioners in the cyber-insurance ecosystem on cyber-security risks and allows for the users to choose their desired level of abstraction in the description of a system. 
\end{abstract}

\section{Introduction} \label{sec:intro}
Cyber-insurance policies provide financial mitigation against defined cyber-risk events up to a specified limit, for which the buyer of insurance pays the insurance provider a premium (fee). The definition of events and calculation of both premia and losses are non-trivial and require significant administrative, financial and legal resources. As such, the areas of cyber-risk and cyber-insurance present a multitude of interesting research problems. 

For many years, there have been attempts in the academic community to organize and structure the research agenda related to cyber-insurance, an early example of which is a prososal by B\"{o}hme and Schwarz (2010)~\cite{bohme2010modelling} for a unifying cyber-insurance modelling framework. Despite considerable subsequent research endeavour, several comprehensive literature reviews~\cite{eling2016we,marotta2017cyber,aziz2020systematic,skeoch2022expanding} suggest that the cyber-insurance research field remains fragmented. This fragmentation motivated a number of eminent researchers and practitioners in the field~\cite{falco2019research} to reiterate the need for a well-defined research agenda for cyber-risk and insurance. One key identified contribution is to quantify cyber-risk and its relationship with organizational structure. 

Insurers will often attempt to assess the security posture of the insurance seeking entity via questionnaire forms (see, for example, Woods et al. (2017)~\cite{woods2017mapping}, and Romanosky et al. (2019)~\cite{romanosky2019content}) and potentially via some limited investigation of their own (for example, perimeter scanning). The use of questionnaires is potentially problematic in cyber-insurance. A company will typically be required to provide answers `to the best of its knowledge'. This creates the risk that (assuming no dishonesty), if a company does not have a complete grasp of its own security posture or maturity, the price of its insurance policy may not correctly capture the risk dynamics of the organization. For example, asking the question `Do you have anti-virus software and/or firewalls?' may seem straightforward. However, there are a number of deficiencies in the question. An answer `yes' does not mean that every machine on the network has anti-virus software installed; it says nothing about the robustness of the firewall configuration or any potential rule exemptions set up. Either of these details could be easily exploited by an attacker to introduce ransomware, for example, into an enterprise network. As such, both the detail of the system architecture and \textit{how} the system is monitored are important.    

\subsection{Problem statement}
\label{subsec:problem-statement}
The key question this paper aims to address is: what is needed for a cyber-insurance assessment to be able to properly capture the intricacies of the system being insured at a level of abstraction that makes the assessment possible to perform on a reasonable timescale yet with sufficient descriptive power? Pricing cyber-insurance policies requires a model to estimate the frequency of attacks and a methodology to estimate the size of losses.\footnote{Insurers usually refer to this as severity, which is the average loss per claim.} Mapping out the detail of an organization's network is one possible starting point, but this only reveals details about the systems an organization uses as opposed to its revenue generating operations, which ultimately determine the scope of possible losses related to a cyber-risk incident. There is therefore a need for an integrated description of the structure of an organization and the IT systems that support delivery of its objectives. Once this relationship is described, the potential scope of losses resulting from cyber-incidents and their probability need to be estimated. Based on these estimates, an insurance company can then quote a premium for reimbursing a customer for these losses should they occur.

This paper proposes a modelling framework drawing on the fields of operations management, security, and economics that can be used to support moving from describing an organization to pricing a cyber-insurance policy. Our motivation is primarily to contribute to the conceptual debate on how to address the significant modelling challenges associated with cyber-insurance pricing and so this paper does not directly calculate premiums for real-world organizations and is not an insurance pricing model. Rather it aims to encourage development of novel modelling approaches to improve the interaction between economic approaches to cyber-insurance pricing and systems assessment. It should be noted that at present, assessment of systemic cyber risk is typically performed using `realistic disaster scenarios' (RDS)~\cite{lloyds-rds,pra2022,eling2022economic}. The modelling framework introduced in this paper has potential to be applied to RDS development, as it has been designed to be both highly flexible and scalable yet operates on a similar level of abstraction to these scenarios.   

\subsection{Related work}
Ruan (2017)~\cite{ruan2017introducing} proposed a framework, \textit{cybernomics}, proposing a combination of methods for estimating the value of digital assets. The work suggests risk units as outputs but has a different focus from the research in our work, as the model of Ruan does not explicitly aim to capture the precise organizational architecture and structure of model organizations. Erola et al. (2022)~\cite{erola2022} present a system that calculates cyber value-at-risk via sequential Monte Carlo simulations. The work includes an example case study based on data provided by an insurance company using risk factors provided by a model vendor. The model proposed in our work might be used to structure a cyber value-at-risk simulation by identifying relevant facets of a target organization to inform simulation parameters and as such the model of Erola et al. and ours might usefully be combined. Calvo and Beltran (2022)~\cite{CALVO2022102612} propose a model for adaptive security controls that is only loosely related to the approach we propose, but may have relevance for insurers who wish to take an active interest in their customers' security posture from a remediative perspective. 

The specific literature on security maturity models relevant to this work is not especially developed. One possible reason for this is that security maturity models are of significant use to managers and practitioners, but are relatively recent and therefore it may be too early to expect an empirical literature to have developed. For completeness, we review a selection of relevant papers here. Mettler (2011)~\cite{mettler2011maturity} provides insight into the development of maturity models from a design science research perspective. Rea-Guaman et al. (2017)~\cite{rea2017comparative} introduce a taxonomy for assessing cybersecurity capability maturity models. Kour et al. (2020)~\cite{kour2020cybersecurity} discuss a cybersecurity maturity model specifically for the railway sector; however, the paper provides a useful general overview of the various CSMM models in existence. Couretas (2018)~\cite{couretas2018introduction} provides a good general introduction to cyber-modelling. Cebula and Young (2010)~\cite{cebula2010taxonomy} propose a taxonomy of operational cyber-risks, organized into actions of people, systems and technology failures, failed internal processes and external events. This may prove useful for the purpose of insurance contract construction.

\subsection{Paper organization}

The remainder of this paper is organized as follows. Section~\ref{sec:model} introduces the structure of our proposed modelling framework --- that is, 
a conceptual tool for constructing models; it is not a model in itself --- without providing precise details of the instantiation of its components. We wish to emphasize that the motivation of this paper is to introduce a \textit{flexible} framework adaptable to a range of modelling applications and that just a small subset of its potential is explored in this introductory paper. Section~\ref{sec:modelling-tools} provides a theoretical overview of the tools that we recommend be used to deploy the modelling framework, namely entity relationship diagrams, security maturity models, and utility functions. Section~\ref{sec:case-studies} applies the modelling framework to three prototypical (i.e. fictional) organizations with a diverse range of security requirements. This provides a demonstration of how the modelling framework might be applied to real world organizations. Finally Section~\ref{sec:discussion} discusses issues pertinent to real world implementation of the model, its advantages and disadvantages, and potential further work. 

\section{Modelling framework overview} \label{sec:model} 
 As discussed in Section~\ref{sec:intro}, the aim of the modelling framework introduced in this paper is to connect a description of an organization's structure and systems with the economic parameters required to price a corresponding cyber-insurance policy via an assessment of the security posture of the organization using maturity models. This approach reconciles system-centric and economic treatments of a cyber-insurance decision problem. It is intended to structure and guide the cyber-insurance assessment and not to act as a universal model for cyber-insurance pricing. 
 
At this introductory stage, we focus purely on introducing the components of the modelling framework and the required parameters rather than characterizing them in detail. The overall aim of the modelling framework is to support making insurance assessments that take account of both the overall structure of an organization and the intricacies of its systems. However, it is not our intention that the precise implementation of the modelling framework should be constrained; different target organizations and/or insurers may require different parameters. Specifically, requirements for the detail of an assessment vary across the insurance industry. Firms writing a large number of small limit policies are far less likely to require a detailed assessment of an insurance customer than a firm writing multi-million dollar limits for a select number of multinational organizations. 

We aim to contribute a methodology for making insurance judgements and a set of parameters that are suitably flexible to be relevant across a range of possible analyses. In due course, full examples of how the modelling framework might be used will be provided. Section~\ref{sec:modelling-tools} will give examples on how our proposed modelling tools might be used to construct insurance policies and Section~\ref{sec:case-studies} will use the modelling framework to deliver case studies for three hypothetical organizations from separate industry groups to demonstrate its functioning and relevance. 

\subsection{Preliminaries}
\subsubsection{Representing organizational structure}
We propose representing organizations using three layers: a management layer, a services layer, and a systems layer. The management layer describes the structure of the organization --- including, but not limited to, what it provides and how it achieves it. The systems layer depicts the infrastructure used in the organization, such as servers, databases or other relevant elements such as manufacturing equipment. The link between the management and systems layers is depicted by service layers. The amount of detail captured by these layers is an important consideration. A description of every communication is likely to be unusable, but a simple overview of management structure too simplistic. The necessary level of detail for an insurance assessment is likely to vary for different types of organizations and insurer risk appetites. While the description we propose might be adequately represented in a table for simple instances, a diagrammatic representation represents a better choice in terms of legibility and ease of interpretation. With an appropriate schematic formulation, a well-constructed diagram might fulfil a diverse range of requirements without needing to be redrawn for different audiences. Entity relationship diagrams (ERDs) are a strong candidate for delivering such a diagrammatic representation. Their notational flexibility allows the user to implement their desired level of abstraction rather than prefiguring an organizational description, which is compatible with the objective of the modelling framework to balance flexibility with descriptive power. ERDs will be fully introduced and explained in Section~\ref{subsec:cyber-insurance-policies}. 

\subsubsection{Describing organizational security posture}
\label{subsec:criticality-sensitivity}
Having established how to represent an organization and its systems, we now require appropriate risk metrics. We propose using criticality and sensitivity as measures for the components of the system that an organizational policymaker may wish to protect. The practice of framing security analyses using the categories of phenomena confidentiality, integrity and availability (often termed the `CIA triad') is commonplace.\footnote{See Pym (2015)~\cite{pymbenthamsgaze} for a discussion.}   
\begin{figure}[ht]
\begin{center}
\begin{tikzpicture}
    \node[entity] (criticality) at (0,0) {Criticality};
    \node[entity] (sensitivity) at (4,0) {Sensitivity};
    \node[entity] (availability) at (-2,-2) {Availability}
        edge(criticality);
    \node[entity] (integrity) at (2,-2) {Integrity}
        edge(criticality) edge(sensitivity);
    \node[entity] (confidentiality) at (6,-2) {Confidentiality}
        edge(sensitivity);
\end{tikzpicture}
\end{center}
\vspace{1mm}
\caption{Relationship between criticality/sensitivity and confidentiality/integrity/availability}
\label{fig:cs-to-cia}
\end{figure}
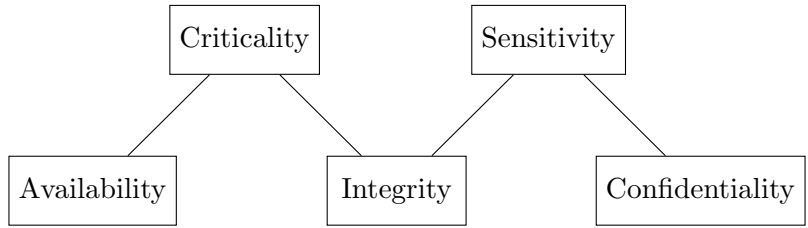
As illustrated in Figure~\ref{fig:cs-to-cia}, confidentiality, integrity and availability can be simplified to criticality and sensitivity. This assumes that availability maps to criticality; confidentiality to sensitivity; and some aspects of integrity map to criticality but others to sensitivity. For the purposes of the organizational assessments in this work, criticality and sensitivity are the simplest possible metrics that permit description of a wide range of different organizational types and architectures while achieving consistency with recognized industry norms. However, a wider range of categories could easily be deployed according to the requirements of the modeller.  

As already outlined, it is important to consider how the structure of an organization and its systems may create risks of some form of cyber-attack and also how that attack might be detected. For a very simple organization, some form of `fit and forget' controls might suffice, but for more complex organizations, the security posture is likely to evolve over time. A framework is accordingly required that is able to capture the notion of change and describe the communications within an organization and externally in a rigorous and structured manner. No single approach is likely to perfectly suit every instance and, consequently, it is probably optimal to settle of a class of models and choose a specific model for the case under consideration. 

The modelling framework introduced in this paper uses \emph{security maturity models} to parametrize the security posture of an organization. The term maturity in this context means how developed the security controls and monitoring are in relation to an organization's processes, objectives and risk tolerance. An analysis of security maturity needs to produce outputs that can be compared in the context of the type of organizational structure, the details of the systems architecture and the characteristics of internal and external communications. One might construct a measure of the openness of an organization and its risk of being attacked or another measure of the desirability of an organization as a target. If correctly developed, a measure of security can be used to predict and analyze attacks and resultant losses, which in turn might be used to aid the construction of probability distributions, which are a key input to actuarial modelling in insurance.    

\subsubsection{Economic modelling}
For the purposes of our modelling framework, an economic model is a function or framework that maps a number of (ideally parsimonious) parameters to an output. There are two clear categories of function of relevance to the challenge at hand: pricing formulae and utility functions. Pricing formulae are a category of functions that, as the name suggests, output the price for a financial instrument based on a number of inputs. Perhaps the most famous example in the finance literature is the Black-Scholes equation,\footnote{See Duffie (1998)~\cite{duffie1998black} for an accessible description.} which outlines the price for an option (the right but not the obligation to buy or sell a security). Utility functions stem from seminal work by von Neumann and Morgenstern\footnote{Fishburn (1989)~\cite{fishburn1989retrospective} provides a concise introduction to the topic.} and are broadly speaking a mathematical description of the preferences of a decision maker. Utility functions require careful formulation and evaluations when used in economic modelling. As this work is intended to outline a conceptual framework for combining systems and economic modelling, we will demonstrate the use of utility functions for model convenience, but it is important to be aware that there is a rich risk management literature (see, for example, Gollier (2001)~\cite{gollier2001} for a starting point) on their use and, for some modelling applications, there may be alternatives to utility functions that better capture the modelled trade-offs. 

Within the field of security economics, Gordon and Loeb (2002)~\cite{gordon2002economics} proposed a model connecting investment in information security and the probability of a security breach. This spawned a derivative field of literature, for example Naldi and Flamini (2017)~\cite{naldi2017calibration}, Gordon et al. (2020)~\cite{gordon2020integrating} and Skeoch (2022)~\cite{skeoch2022expanding}. Whilst the Gordon-Loeb model is attractive in its tractability and simplicity, the choice of parameters is at best subjective and at worst contentious. Consequently, rigour, care and justification are required for an implementation of the model to have practical use. The usual technique in economics for populating model parameters is to use econometrics, which provides mathematical tools for deriving parameters from data in a broad sense. However, this is difficult for cyber-risk problems as there is a paucity of publicly available data and that available is other sparse or has a very short history. This is a significant (and at times valid) criticism of the use of economics in security and it is hoped that the contributions of this research illustrate a potential alternative means of deriving useful insights from economic models. 

\subsection{Modelling framework structure and parameters}
\label{subsec:model-structure-params}
\begin{figure}[ht]
	\centering
        \resizebox{\textwidth}{!}{%
	\includegraphics{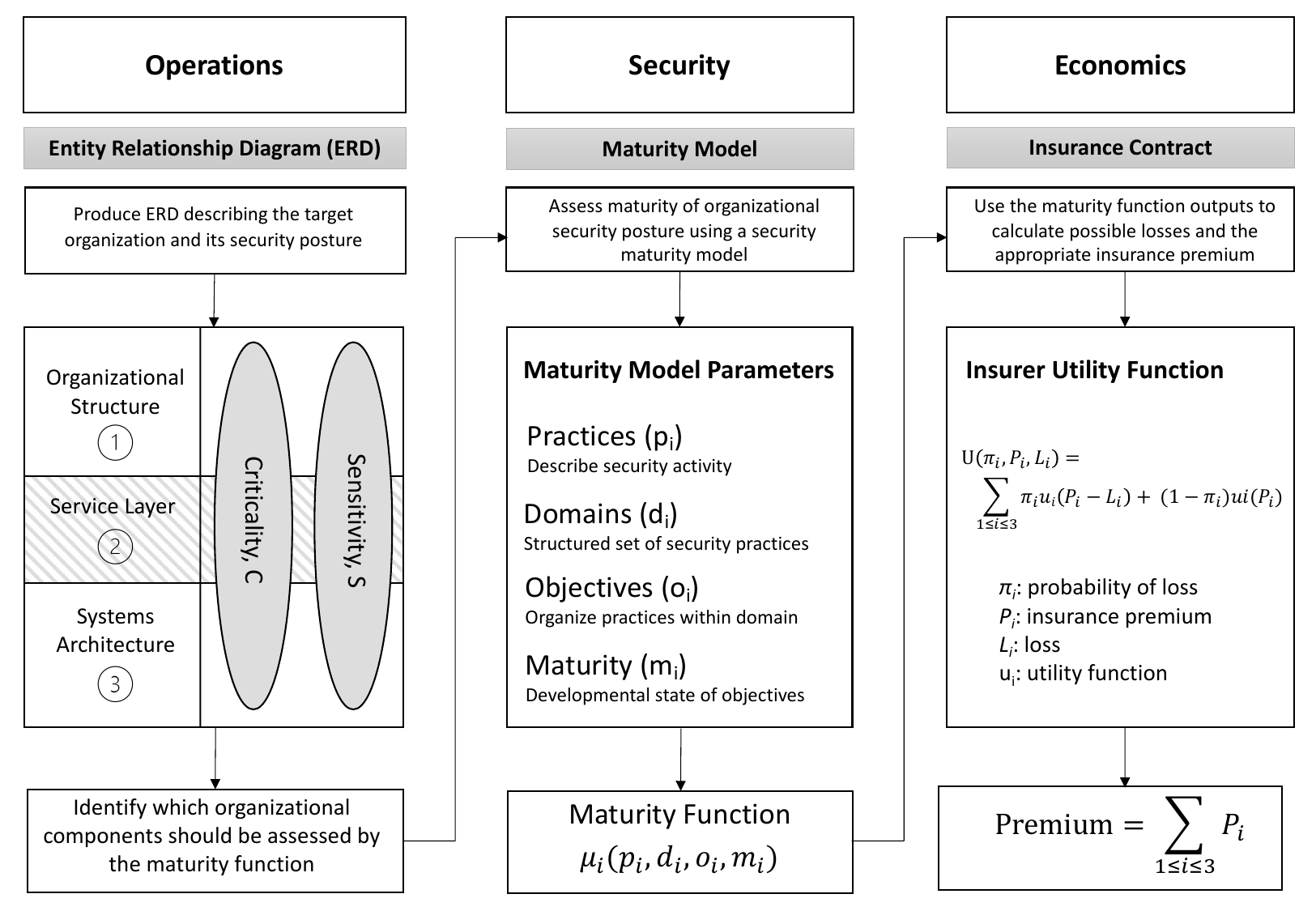}
        }
	\caption{Conceptual overview of model}
	\label{fig:paper-overview}
\end{figure}

Figure \ref{fig:paper-overview} outlines the structure of the modelling framework. Table~\ref{tab:appendix-maths-symbols} in Appendix~\ref{appendix:mathematical-symbols} details the mathematical notation used in this paper for convenience. The description of the maturity model draws on the models presented in Table \ref{table:maturity-models} and described in Section~\ref{sec:sec-mat-models}. We wish to emphasise that the form of the security maturity model shown in the specification is not intended to be prescriptive. At these stage, we simply introduce the parameters that comprise the proposed modelling framework and to retain flexibility do not constraining their form and range of values. The aim of the modelling framework is to illustrate how models from different disciplines might usefully be combined to improve cyber-insurance pricing and not to dictate the best way to describe organizational security posture. 
\begin{itemize}
\item The parameters $C_i$ and $S_i$, for $1 \leq i \leq 3$, represent the components of the operations 
architecture --- organizational structure (1), service layer (2), and systems architecture (3) --- that deliver 
respectively those aspects of the operations for which criticality and sensitivity are required by 
the policy-maker to be protected.  
\item The parameters $p_i$, $d_i$, $o_i$, $m_i$ denote respectively representations of the effectiveness 
in protecting $C_i$ and $S_i$ of the practices, domains, objectives, and maturity of the layers of 
the architecture as assessed by the chosen maturity model.  
\item The key aspects of this system dependency are the following: the interdependencies between the system's layers and the interdependencies between the individual components of the system. These interdependencies are represented explicitly, through relationships and attributes, in the Entity Relationship Diagrams~\cite{chen1976entity} that we 
describe in Section~\ref{subsec:ERDs}. Our proposed use of Entity Relationship Diagrams in this context is illustrated in the case studies that we present in Section~\ref{sec:case-studies}. 
\item The maturity function $\mu_i(p_i,o_i,d_i,m_i)$ transforms the parameters of the chosen maturity model into parameters to be input into an economic model. The form and outputs of the maturity function depend on both the chosen maturity model and the target economic model. The maturity model may be based on or aligned with accepted security standards such as ISO27001, though this is not a requirement.
\item A condition on the formulation of our model is that the analysis of the system's security posture that is provided by the maturity model must support the instantiation of the required utility function for the insurance contract. 
\item The loss-generating function, $L_i(\mu_i,C_i,S_i)$, specifies the monetary losses that may be incurred by a deviations in $C_i$ and/or $S_i$ from their values in the intended operating state specified by the policymaker. The mapping of these deviations to losses is determined by $\mu_i$. 
\item The probability of a loss occurring due to a deviation in $C_i$ and/or $S_i$ is denoted by $\pi_i \in [0,1]$. The calculation of the probability may be a function purely of the properties and/or maturity of the operations architecture component itself (endogenous factors) or take into account external influences or observations (exogenous factors).   
\item $P_i(\pi_i,L_i)$ are the components of the premium charged by the insurer from its assessment of each component of the operations architecture. The premium is a function of loss, $L_i$ and $\pi_i$, the probability of that loss occurring. The final premium charged for the contract is $P = \sum\limits_{1 \leq i \leq 3}  P_i$.
\item The utility function for the insurer is defined as \[U(\pi_i,P_i,L_i) = \sum\limits_{1 \leq i \leq 3} \pi_i u_i(P_i - L_i) + (1 - \pi_i)u_i(P_i)\]The objective of an economically rational insurer is to set premia that maximize this utility function. $u_i$ is a utility function that describes the risk preferences of the insurer.   
\end{itemize}

\section{Setting up the Modelling Framework} \label{sec:modelling-tools}

This section provides definitions and examples of the tools deployed in the modelling framework following the informal introduction of the relevant ideas in the previous section. The modelling framework uses three tools drawn from the fields of Operations, Security, and Economics:           

\begin{framed}
\begin{itemize}[itemindent=4em]
    \item[\textbf{Operations}] \emph{Entity Relationship Diagrams} describe the organization and its systems 
    \item[\textbf{Security}] \emph{Maturity Models} assess the security posture of the organization 
    \item[\textbf{Economics}] \emph{Utility Functions} price the insurance contract
\end{itemize}
\end{framed}
Before considering each of the three tools in detail, we briefly describe how they fit together to deliver the modelling framework. Entity relationship diagrams provide a description of three elements of the organization: its organizational structure, its systems architecture and the service layer that connects the two aforementioned elements. The complete entity relationship diagram may then be used to ascertain which of its components are susceptible to degradation in either criticality or sensitivity or both. With this description established, a maturity model is then used to describe the security posture of the organization. The maturity model is used to generate a maturity function, $\mu_i(p_i, d_i, o_i, m_i)$. The parameters practices ($p_i$) and domains ($d_i$) are provided by the maturity model. Objectives ($o_i$) are determined by a policy-maker. In simple cases, the insurer might be the sole policy-maker but for more complex organizations, high-level objectives from the insurer may be more precisely implemented by a policy-maker within the organization (for example, a CISO\footnote{Chief Information Security Officer}). The maturity $m_i$ specifies how well the objectives are currently met by the security practices of the organization. 

The set of elements susceptible to degradation in criticality or sensitivity are potentially loss-generative in a monetary sense. In order to price an insurance policy, an insurer needs an expected loss for the policy and a probability of loss for that policy. In the model proposed here, the insurer uses a utility function to price the premium. The maturity function provides the parameters needed to estimate probabilities ($\pi_i$) and losses ($L_i$) for each element of the organizational structure, which then generate a respective premium for each element. The sum of these premia then gives the overall premium for the policy. As an insurer gains experience of losses over time, it will likely gain insight into the relationship between practices and the parameters $\pi_i$ and $L_i$ and thus refine the objectives required to provide insurance cover or alter its pricing.    

It should be noted that reimbursement of a loss by an insurance company is not automatic. The holder of the insurance policy must first present a claim to the insurance company. The insurer then assesses the validity of the claim, and how much of it to reimburse. Individuals fulfilling this function are known as loss-adjusters. 

\subsection{Entity Relationship Diagrams} \label{subsec:ERDs}
\subsubsection{Overview} \label{subsubsec:erd-overview}

A description of organizational structure and systems architecture requires a language that is able to express systematically the key components of a structure and their relationships. There is a rich literature on formal systems modelling, which is too large and diverse for us to review fully within the scope of this paper. Examples of this discipline include the strongly compositional approach to systems modelling (e.g., among many,  \cite{Birtwistle1979,Milner2009,CMP2012,caulfield2015improving}), UML~\cite{rumbaugh-uml, breu1997towards, elmasri2000fundamentals}, and system dynamics~\cite{forrester1997industrial, sterman2002system}.   

In the modelling framework proposed in this research, we contend that the language of entity relationship diagrams has just the right level of conceptual analysis, abstraction, and descriptive power to sufficiently characterize the structure of an organization and its systems so as to support an analysis of its security posture for the purposes of an insurance assessment. The formal logic associated with the strongly compositional approach to systems modelling (see, for example, \cite{CMP2012}) is not required for the purposes of insurance assessments as insurance is always uncertain and therefore the modelling task is to \emph{estimate} to the best of the modeller's ability rather than aiming to \emph{prove} the properties of a system. The distinction between ERDs and UML is somewhat more subtle; UML is a language for creating diagrams whereas ERDs are a type of diagram. The nature of the modelling task for assessing organization structure relationships is a subjective task and consequently, the use of ERDs is more appropriate for our use case. System dynamics provides a methodology for rationalizing the behaviour of complex systems; it has a clear potential application to network modelling but is not particularly helpful for attempting to describe a diverse range of systems using a common set of models and parameters.  

The ERD language has a grammar, originally introduced by Chen(1976)~\cite{chen1976entity} for representing data structure, which describes the connectivity and properties of the components of the language. 
\begin{itemize}
    \item \emph{Entities} are `things', which can be distinctly identified. These might, for example, be organizational departments or assets, network resources, or control devices such as firewalls or intrusion detection systems (IDSs). 
    Entities may be either informational or physical. 
    \item \emph{Relationships} describe associations between entities. 
    \item \emph{Attributes} are associated with both entities and relationships. They are parameters of the entity or relationship they describe.
    \item A system is described by connecting entities, relationships, and attributes in a graph. The representation of the connection may describe properties of that connection, such as cardinality or modality. 
\end{itemize}
The description of a system via this language is known as an entity relationship diagram (hereafter referred to as an ERD). The components of an ERD map to the grammatical structure of natural language as depicted in Table~\ref{table:erd-nat-lang}. 

\begin{table}[!htb]
    \centering
    \begin{tabular}{|c|c|}
        \hline ERD & Natural Language  \\
        \hline
           Entity & Noun \\
           Relationship & Verb \\
           Attribute (Entity) & Adjective \\
           Attribute (Relationship) & Adverb \\
           \hline 
    \end{tabular}
    \vspace{1mm}
    \caption{Mapping of ERD and natural language grammar}
    \label{table:erd-nat-lang}
\end{table}

Properties of the connectivity between the different elements of an ERD, such as \textit{cardinality} (whether the relationship is one-to-one, one-to-many, many-to-one, or many-to-many) and \textit{modality} (whether the relationship is mandatory or optional). Of course, other logical relationships might be established and depicted as required.

ERD notation was originally developed as a model for data. A key motivation for its genesis was to reconcile different models for data (see Chen (1976)~\cite{chen1976entity} for a detailed discussion and references). In this paper, we aim to demonstrate that there is significant potential in using ERDs to describe complex systems more generally with great flexibility and compatibility with a range of logical requirements. For example, in formal systems modelling, combining systems diagrams is known as \emph{composition}. ERDs deliver this by being readily combinable using appropriate combinators (such as lines or arrows) to link separate diagrams. This is a way of delivering what, in common language, is known as scalability.        

\subsubsection{ERD Notation}
\label{subsubsec:notation}

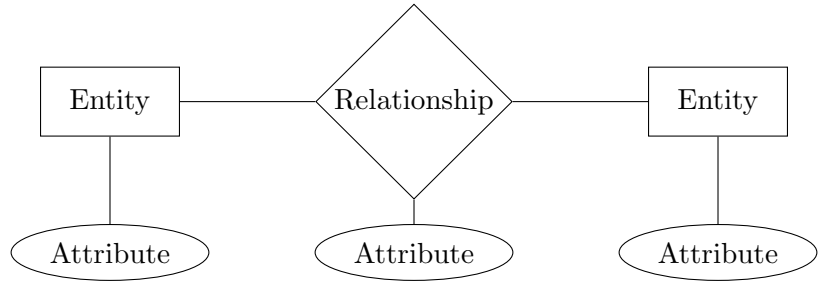
\begin{figure}[!htb]
\begin{center}
\begin{tikzpicture}
    \node[entity] (entity) {Entity};
    \node[entity] (entity2) at (8,0) {Entity};
    \node[relationship] (relationship) at (4,0) {Relationship}
        edge(entity)
        edge(entity2);
    \node[attribute] (attribute1) at (0,-2) {Attribute}
        edge(entity);
    \node[attribute] (attribute3) at (4,-2) {Attribute}
        edge(relationship);
    \node[attribute] (attribute2) at (8, -2) {Attribute}
        edge(entity2);
\end{tikzpicture}
\end{center}
\caption{A simple entity relationship diagram}
\label{fig:erd-example}
\end{figure}

Having established the motivation for deplying ERDs in this work, we now introduce their notation. The examples in this work are limited to entities with optional attributes, relationships without attributes, and lines to represent connectivity (Figure~\ref{fig:erd-example}). 

This provides the sufficient level of abstraction to apply the model introduced by this work and exemplified by the case studies in Section~\ref{sec:case-studies}. However, we wish to re-emphasize that the level of abstraction is a \emph{choice} and not a \emph{prefiguration}. Depending of the desired intricacy of a particular ERD, more detailed notation might be required. This work makes use of Chen notation~\cite{chen1976entity} to describe the components of a system. The foundations of this notation have already been outlined in Section~\ref{subsubsec:erd-overview} and for completeness we now describe the notation in full. 

\begin{itemize}
    \item \emph{Entities} are something distinctly identifiable, such as data, an organizational department, a server, a firewall. They are the fundamental building blocks of the ERD language. 
    \item \emph{Weak Entities} are entities that cannot be identified by their attributes alone; they require another entity to constitute a unique identifier. A simple example is a room, which cannot exist without a building. 
    \item \emph{Associative Entities}\footnote{See \cite{tupper2011data} for a careful explanation.} are entities that resolve many-to-many relationships. For example, in a database of student --- class relationships, one could construct an enrollment entity with a teach relationship to a teacher and which connects students to classes.  
    \item \emph{Attributes} are properties of entities or relationships. For example, in an organization, the Finance function might have the attribute `department'.
    \item \emph{Key Attributes} are attributes which uniquely identify an entity, for example, an employee identification number.
    \item \emph{Derived Attributes} are calculated from other attributes. For example, an employee's age will increase over time but their date of birth remains constant. Age is therefore a derived attribute.
    \item \emph{Multi-valued Attributes} are attributes that may have more than one value associated with the key.
    \item \emph{Relationships} describe how entities interact.
    \item \emph{Weak Relationships} are non-identifying relationships; formally the primary key of one of the related entities does not contain a primary key component of the other related entities. 
\end{itemize}
The diagrammatic representation of the above components is provided in Figure~\ref{fig:chen-notation} in Appendix~\ref{appendix:erd-notation}. Chen used a solid line to indicate a mandatory relationship between components and a dashed line to indicate optional relationships. A richer set of combinators are provided by Barker~\cite{barker1990case} notation (Figure~\ref{fig:barker-notation}) and Crow's foot notation~\cite{everest1976basic}(Figure~\ref{fig:crows-foot-notation}). Both notations allow for cardinality (number of relationships) and modality (whether the relationship is optional) to be fully conveyed. Chen notation only allows for a binary description of modality between two components.     

\subsubsection{Example ERD}

Figure~\ref{fig:payroll-erd-with-c-and-s} shows how the payroll function of an organization could be represented using an ERD. This example has been constructed to illustrate the variety of Chen notation that could be used to describe a system. It is intended as a stylized example rather than claiming to be an optimal representation of a payroll function. In the example, there are two organizational departments, Finance and Human Resources, which are entities with an attribute denoting that they are departments. The finance department is responsible for paying employees, which are an entity with a multi-valued attribute `resource' to recognize that employees may have multiple attributes. There is a relationship `pays' between the finance department and the employees. There is a payroll database, which is a a weak entity with the attribute database; this database contains the entity records with attribute data. It is classified as a weak entity here as it exists to fulfill a purpose and without the payroll function, it would not exist. The contents of the database are delivered via an associative entity, `records', within the service layer. The service layer in the model is a natural location for associative entities depending on the nature of the organization or entity being depicted as the service layer connects the management and systems layers of the organization in the model. The records have the keyed attribute, `Data', as each employee record has a unique identifier. The shading in the diagram denotes which elements are critical and which are sensitive; the use of this will be explained in detail in Section~\ref{subsubsec:example-payroll-smm-assessment}.

\usetikzlibrary{er, positioning}
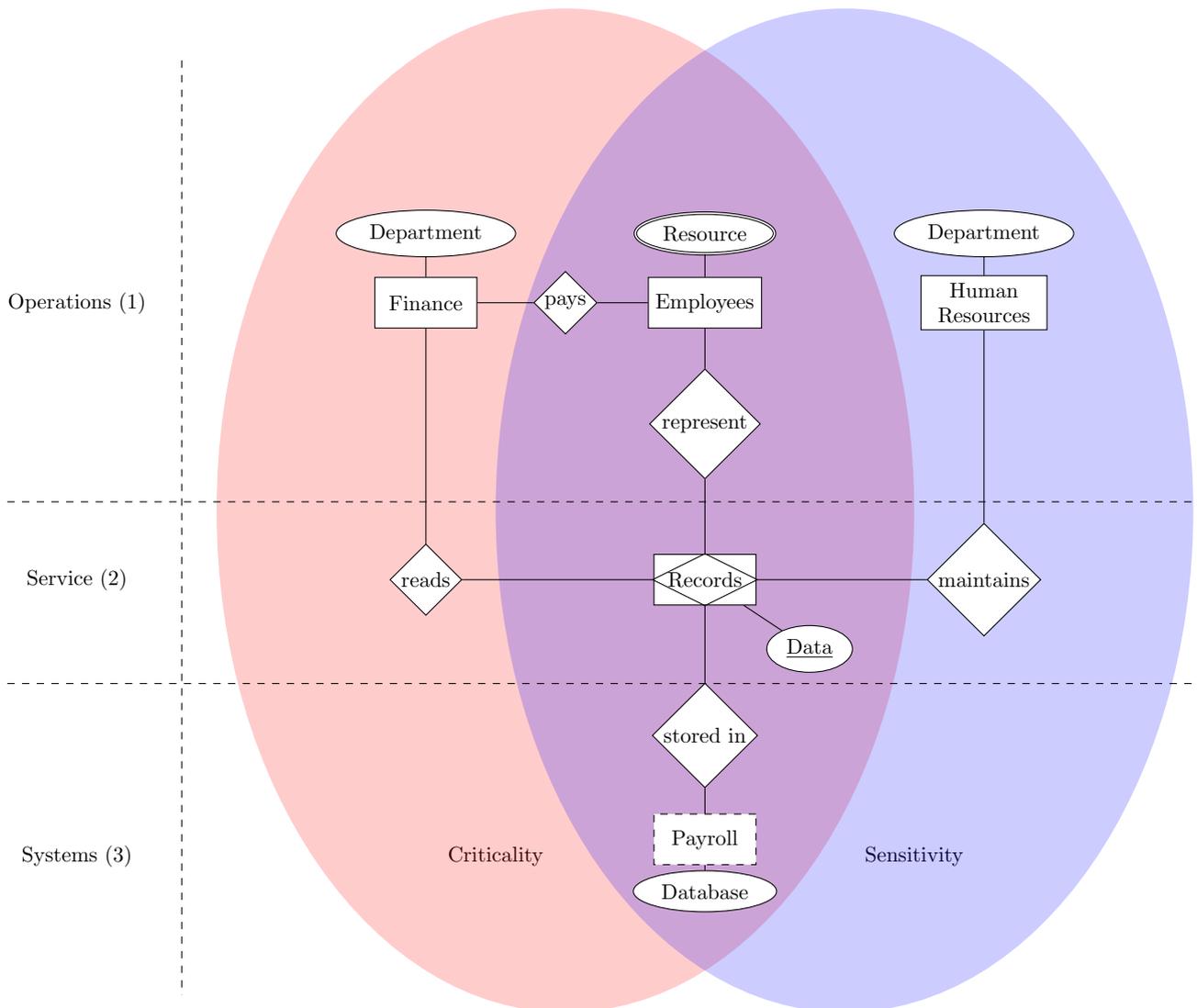
\begin{figure}[ht]
\begin{center}
\begin{tikzpicture}
       [every node/.style={fill=white, scale=0.8}]
     \payrollerd
    \begin{scope} [on background layer]
    \node[] (Criticality) at (1, -10) {Criticality};
    \node[] (Sensitivity) at (7, -10) {Sensitivity};
    \begin{scope} [fill opacity = 0.2]
    \fill[red] (2,-5) ellipse (5 and 7.25);
    \fill[blue] (6, -5) ellipse (5 and 7.25);
    
    \end{scope}
    \end{scope}
\end{tikzpicture}
\end{center}
\caption{Payroll model with criticality and sensitivity}
\label{fig:payroll-erd-with-c-and-s}
\end{figure}
The role of entity relationship diagrams within the model is to 
\begin{itemize}
    \item describe the regions or zones for which different maturities can be assigned, 
    \item show for which of these regions criticality and sensitivity are required to be protected, and 
    \item represent the connectivity between different elements of the organizational structure.
\end{itemize}
The entity relationship diagrams generated in the organizational assessment could also potentially be used to identify \emph{vulnerabilities} arising from connectivity within the organization and its systems so that an assessment of controls can be undertaken. 

We contend that our choice of ERDs to represent the structure of the system is at the right level of abstraction to be able to capture the different security postures in different regions of the architecture without introducing complexity or detail that is not needed for this purpose. Furthermore, we observe that, if required, ERDs can be used to give more detailed descriptions of specific components of the description of an organization. This can be done in a manner that is compositional, with a more complex model of a component replacing a simpler one (or vice versa). Moreover, we would suggest that our choice of such a minimal representation of the architecture of the underlying system supports the scaling of our methodology to larger, more complex systems. 

\subsection{Security Maturity Models} \label{sec:sec-mat-models}

 Security Maturity Models (SMMs) are a tool used by organizations to describe their security posture in a structured manner. They achieve this by stipulating a framework on how to group processes and resources related to cyber-security and then assess how developed, or mature, is the posture of the organization. SMMs are not just a tool for self-analysis; they are gaining increasing prominence as a reference for the minimum standards required by external contractors to sensitive organizations, such as government or the military. There is no one standard SMM in current use. Table~\ref{table:maturity-models} lists the SMMs that appear to be in common use based on web searches. While we believe it to be comprehensive, it is not intended to be exhaustive.     
\begin{table}[!htb]
\begin{center}
	\begin{filecontents*}{smm-summary.csv}
Model,Organization,URL
Personnel Security Maturity Model,UK Centre for the Protection of National Infrastructure,https://www.cpni.gov.uk/personnel-security-maturity-model
IoT Security Maturity Model,Industrial Internet Consortium,https://www.iiconsortium.org/pdf/SMM_Description_and_Intended_Use_V1.2.pdf
IA Maturity Model (HMG),NCSC,https://www.ncsc.gov.uk/information/hmg-ia-maturity-model-iamm
Cybersecurity Maturity Model Certification,US Department of Defense,https://www.acq.osd.mil/cmmc/
Community Cyber Security Maturity Model,UTSA CIAS,https://cias.utsa.edu/research/maturity-model/
ICS-SCADA,ENISA,https://www.enisa.europa.eu/publications/maturity-levels/at_download/fullReport
C2M2,US DOE,https://c2m2.doe.gov/
\end{filecontents*}
\pgfplotstabletypeset[
	col sep = comma,
	columns = {Model,Organization},
	columns/Model/.style={verb string type, column name={Model}, column type={|l} },
	columns/Organization/.style={verb string type, column name={Authoring organization}, column type={|l|} },
	every head row/.style={before row=\hline,after row=\hline},
	every last row/.style={after row=\hline},
	]{smm-summary.csv}
	\vspace{1mm}
	\caption{Currently used security maturity models}
	\label{table:maturity-models}
	\end{center}
\end{table}
The exact specification and terminology vary from model to model, but most follow the basic structure of:
\begin{itemize}
    \item \emph{Practices} ($p_i$) describe single security activities and are \textit{how} maturity is achieved 
    \item \emph{Domains} ($d_i$) provide a structured set of security practices, grouped by area.  
    \item \emph{Objectives} ($o_i$) are \textit{what} the organization must do for a practice to be deemed as met.
    \item \emph{Maturity} ($m_i$) states the developmental state of objectives, usually assigning a level from a predetermined set. 
\end{itemize}
Any of these elements may be aligned fully or partially with existing cybersecurity standards. The preeminent global standard is ISO27001, which comprises controls organised by category. Within the maturity model, practices are equivalent to controls in the standard and domains are the same as categories. We assume here that the impact on premia of the degree of compliance with standards such as ISO27001 derives from the effect of the level 
of compliance on the parameters mentioned in the maturity model (Figure~\ref{fig:paper-overview} and above).  

Other examples of cybersecurity standards include the NIST Cybersecurity Framework in the USA, Cyber Essentials in the UK, the Essential Eight in Australia and the BSI IT-Grundschutz in Germany. Cyber-security standards generally need to have broad applicability and coverage whereas maturity models may be tailored to a specific use case. The maturity model allows for the same structure and rigour as the cyber-security standards but may place greater emphasis on certain elements of security. 

To illustrate how this works in practice, we use an example from the CMMC.\footnote{A good resource for the CMMC is CyberAssist (https://ndisac.org/dibscc/cyberassist), provided by the US National Defense Information Security and Analysis Center.} A practice might be to disconnect a user after n minutes of inactivity, which is AC.L2-3.1.11 Session Termination in the CMMC. The practice is part of the Access Control (AC), one of 14 domains within the CMMC, which are aligned with the NIST standard SP 800-171. An objective to which this practice might belong could be `Ensure unintended users do not gain access to the system', which might include other practices from the AC domain relevant to the organizational system. Within the CMMC, to attain AC Maturity Level 1,\footnote{The CMMC is, at the time of writing, transitioning from v1 to v2 with the number of maturity levels reducing to 3. This example uses the v2 specification.} the practices Authorized Access Control, Transaction \& Function Control, External Connections, and Control Public Information are required to be implemented. A further 18 incremental practices are required to achieve Level 2 for this domain. 

\subsubsection{Security maturity model effectiveness parameters}

In Section~\ref{subsec:model-structure-params}, we stated that the security maturity model is used to produce a maturity function, $\mu_i$, which takes as parameters $p_i$, $d_i$, $o_i$, and $m_i$ which denote respectively representations of the effectiveness in protecting criticality and sensitivity of the practices, domains, objectives, and maturity of the components of the system architecture as assessed by the chosen maturity model. At that juncture, we refrained from giving a general characterization. We now explain how the parameters \emph{might} be populated for the purposes of assessing payroll function in Figure \ref{fig:payroll-erd-with-c-and-s} and the case studies that follow in Section \ref{sec:case-studies}.  The precise and rigorous population of the effectiveness parameters for a maturity model is a significant and detailed piece of work and is intended to be undertaken by a specialist team within an organization (or, possibly, external consultants). 
\begin{itemize}
    \item $p_i$ is a vector with elements taking values across $[-1,1]$. The elements represent whether each practice in the maturity model is not met (-1), not relevant (0) or met (1). 
    \item $d_i$ is a vector with elements taking values across $[0,1]$ representing the importance of each domain to the insurer or policymaker for their objectives.
    \item $o_i$ is a vector with elements taking values across $[0,1]$ representing how well the assessed organization meets the stipulated objectives.
    \item $m_i$ states the overall maturity level achieved by the organization across $[0,1]$. Different maturity models have different level structures; thus, $m_i=0$ denotes complete immaturity and $m_i=1$ denotes complete maturity.
\end{itemize}
These parameters would allow an insurer selling a specific insurance policy to quickly evaluate the maturity level of the security of organizations seeking insurance in a robust and systematic manner. An insurer might then compile a table or matrix of premium rates compared with maturity to allow automatic policy pricing. 

\subsubsection{Example security maturity assessment of payroll function}
\label{subsubsec:example-payroll-smm-assessment}
In the payroll function depicted in Figure~\ref{fig:payroll-erd-with-c-and-s}, there are three entities in the Management layer: Finance, Employees and Human Resources. The Finance Department is defined as critical as without money being directed as required, the organization cannot function. Human Resources is defined as sensitive, as it is responsible for confidential data and maintaining its integrity. The service layer compromises records, which are read by Finance and maintained by HR. The records are stored in the payroll database, which resides within the systems layer. 

\begin{table}[ht]
    \begin{center}
    \begin{tabular}{|r|p{5cm}|p{5cm}|}
    \hline
    & \textbf{Criticality} & \textbf{Sensitivity} \\ \hline
    \textbf{Management} &(C1) Transactions must be processed when required and recorded & (S1) Employee data must be kept integral and confidential \\ \hline
    \textbf{Service} &(C2) Records must be available &(S2) Records must only be accessed or modified by approved persons \\ \hline
    \textbf{Systems} &(C3) The database must be available except for scheduled maintenance and must be integral &(S3) The database must have appropriate information security controls \\ \hline
    \end{tabular}
    \end{center}
\vspace{1mm}
\caption{Objectives for payroll example}
\label{table:payroll-example-objectives}
\end{table}

Table~\ref{table:payroll-example-objectives} states objectives for the payroll function that might be determined by an organizational policymaker. We use the notation C and S to number those objectives that relate to criticality and sensitivity, respectively. As noted by Romanosky (2019)~\cite{romanosky2019content}, insurance companies tend to use some standardized premium rates adjusted for various factors, which are typically multiplicative. In the initial example payroll example, deviations from the objectives outlined in Table~\ref{table:payroll-example-objectives} with respect to criticality and sensitivity might incur losses. However, these might be mitigated by adopting security \emph{practices} stated in the maturity model. First, each of the objectives needs to be mapped to \emph{domains} from the maturity model (in this case, the CMMC). Table~\ref{table:payroll-domain-objectives-matrix} shows one possible choice of $d_i$ for the payroll example. 

\begin{table}[!htb]
\begin{center}
    
	\begin{filecontents*}{payroll-domain-objective-matrix.csv}
Domain,C1,C2,C3,S1,S2,S3
Access Control (AC),0,0,0,0,1,1
Audit and Accountability (AU),1,0,0,1,0,0
Awareness and Training (AT),1,0,0,1,0,0
Configuration Management (CM),0,1,0,0,1,1
Identification and Authentication (IA),0,0,0,0,1,1
Incident Response (IR),1,1,0,0,0,0
Maintenance (MA),0,0,1,0,0,0
Media Protection (MP),0,0,0,0,0,1
Personnel Security (PS),0,0,0,0,1,0
Physical Protection (PE),0,0,1,0,0,1
Risk Assessment (RA),1,0,0,0,0,0
Security Assessment (CA),0,0,0,1,0,0
System and Communications Protection (SC),1,0,0,0,0,1
System and Information Integrity (SI),1,0,1,0,1,1
\end{filecontents*}
\pgfplotstabletypeset[
	col sep = comma,
	columns = {Domain,C1,C2,C3,S1,S2,S3},
	columns/Domain/.style={verb string type, column name={Domain}, column type={|l|}},
	columns/C1/.style={postproc cell content/.code={
	    \ifnum##1=1 \pgfkeysalso{@cell content=\cellcolor{lightgray}{##1}}\fi}},
	columns/C2/.style={postproc cell content/.code={
	    \ifnum##1=1 \pgfkeysalso{@cell content=\cellcolor{lightgray}{##1}}\fi}},
	columns/C3/.style={postproc cell content/.code={
	    \ifnum##1=1 \pgfkeysalso{@cell content=\cellcolor{lightgray}{##1}}\fi}, column type={c|}},
	columns/S1/.style={postproc cell content/.code={
	    \ifnum##1=1 \pgfkeysalso{@cell content=\cellcolor{lightgray}{##1}}\fi}},
	columns/S2/.style={postproc cell content/.code={
	    \ifnum##1=1 \pgfkeysalso{@cell content=\cellcolor{lightgray}{##1}}\fi}},
	columns/S3/.style={column type ={c|},postproc cell content/.code={
	    \ifnum##1=1 \pgfkeysalso{@cell content=\cellcolor{lightgray}{##1}}\fi}},
	every head row/.style={before row=\hline,after row=\hline},
	every last row/.style={after row=\hline}
	]{payroll-domain-objective-matrix.csv}
    \vspace{1mm}
	\caption{$d_i$ parameters for payroll example}
	\label{table:payroll-domain-objectives-matrix}
	\end{center}
\end{table}
The next step in the procedure for maturity assessment is to populate the matrix of practices, $p_i$, according to which practices are met. This is clearly highly dependent on the precise organizational policies and systems implementation and is consequently difficult to stylize in a meaningful way for the payroll example. However, as an illustration of how this might be performed, Table~\ref{table:cmmc-practices-count} summarizes the number of practices required for each domain within the maturity model to achieve either Level 1 or Level 2 within the maturity model. At time of writing, the required practices to achieve Level 3 within the CMMC had not yet been made publicly available. 

For the purposes of the simple payroll example, we limit the scope of the assessment to the 17 practices that comprise Level 1 within the CMMC (Table~\ref{table:payroll-practices-matrix}). The reader interested in the detailed description of the practices is referred to the CMMC Model Overview\cite{cmmc-model-overview}. Table~\ref{table:payroll-practices-matrix} should be read as the necessary set of practices an organization should meet to achieve an insurance premium pricing of Level 1 in this example. Some practices apply only to certain layers; for example, AC.L1-3.1.22, details external connections, which would be specified at the service layer (2) within the organizational model. The extent to which an organization meets these practices in the assessment of its maturity will determine how much of a discount it would receive relative to the baseline premium charged by an insurance company.   

\begin{table}[!htb]
\begin{center}
    \begin{filecontents*}{cmmc-practices-count.csv}
Domain,Level 1,Level 2,Total
Access Control (AC),4,18,22
Audit and Accountability (AU),0,9,9
Awareness and Training (AT),0,3,3
Configuration Management (CM),0,9,9
Identification and Authentication (IA),2,9,11
Incident Response (IR),0,3,3
Maintenance (MA),0,6,6
Media Protection (MP),1,8,9
Personnel Security (PS),0,2,2
Physical Protection (PE),4,2,6
Risk Assessment (RA),0,3,3
Security Assessment (CA),0,4,4
System and Communications Protection (SC),2,14,16
System and Information Integrity (SI),4,3,7
Total,17,93,110
\end{filecontents*}
\pgfplotstabletypeset[
    col sep = comma,
    columns = {Domain,Level 1, Level 2,Total},
    columns/Domain/.style={verb string type, column name={Domain}, column type={|l|}},
    columns/Total/.style={column type = {|c|}},
    	every head row/.style={before row=\hline,after row=\hline},
	every last row/.style={before row=\hline,after row=\hline}
    ]{cmmc-practices-count.csv}
    \vspace{1mm}
    \caption{Number of practices for each Domain in the CMMC by Level}
    \label{table:cmmc-practices-count}
\end{center}
\end{table}

\begin{table}[!htb]
\begin{center}
	\begin{filecontents*}{cmmc-l1-practices.csv}
ID,Practice,p1,p2,p3
AC.L1-3.1.1,Authorized Access Control,1,1,1
AC.L1-3.1.2,Transaction and Function Control,0,1,1
AC.L1-3.1.20,External Connections,0,1,0
AC.L1-3.1.22,Control Public Information,0,0,0
IA.L1-3.5.1,Identification,0,1,1
IA.L1-3.5.2,Authentication,1,1,1
MP.L1-3.8.3,Media Disposal,1,0,0
PE.L1-3.10.1,Limit Physical Access,1,0,0
PE.L1-3.10.3,Escort Visitors,1,0,0
PE.L1-3.10.4,Physical Access Logs,1,1,0
PE.L1-3.10.5,Manage Physical Access,1,0,0
SC.L1-3.13.1,Boundary Protection,1,1,1
SC.L1-3.13.5,Public-Access System Separation,0,0,0
SI.L1-3.14.1,Flaw Remediation,1,0,0
SI.L1-3.14.2,Malicious Code Protection,0,0,1
SI.L1-3.14.4,Update Malicious Code Protection,0,1,1
SI.L1-3.14.5,System and File Scanning,0,1,1
\end{filecontents*}
\pgfplotstabletypeset[
	col sep = comma,
	columns = {ID,Practice,p1,p2,p3},
	columns/ID/.style={verb string type, column name={CMMC Practice}, column type={|l|}},
        columns/Practice/.style={verb string type, column name={Description}, column type={|l|}},
	columns/p1/.style={column name={$p_1$},postproc cell content/.code={
	    \ifnum##1=1 \pgfkeysalso{@cell content=\cellcolor{lightgray}{##1}}\fi}},
	columns/p2/.style={column name={$p_2$},postproc cell content/.code={
	    \ifnum##1=1 \pgfkeysalso{@cell content=\cellcolor{lightgray}{##1}}\fi}},
	columns/p3/.style={column name={$p_3$},postproc cell content/.code={
	    \ifnum##1=1 \pgfkeysalso{@cell content=\cellcolor{lightgray}{##1}}\fi}, column type={c|}},
	every head row/.style={before row=\hline,after row=\hline},
	every last row/.style={after row=\hline}
	]{cmmc-l1-practices.csv}
    \vspace{1mm}
	\caption{$p_i$ parameters for payroll example}
	\label{table:payroll-practices-matrix}
	\end{center}
\end{table}
\subsection{Pricing cyber-insurance using utility functions}
\label{subsec:cyber-insurance-policies}
For the purposes of this model, a cyber-insurance policy is a contract that provides reimbursement for losses, $L$ at a cost of premium, $P$, which may be expressed as a percentage premium rate, $p$. Usually, an insurance buyer will not be able to buy unlimited insurance but will be offered \emph{cover} up to a certain amount. However, the amount of cover an insurer provides represents a limit on its potential cash premium income and the optimal maximum limit offered will accordingly be carefully assessed by the insurer. In this model, we make the simplifying assumption that for the purpose of pricing premia, cover is \emph{ex ante} equal to potential losses. The decision of the insurer is framed in terms of utility maximization in the sense established by von Neumann and Morgenstern (1947)~\cite{vonNeumannMorgenstern}. The utility function, $u_i$, describes the risk preferences of the insurance company (in this case) and may take a number of forms. A commonly used utility function describes constant absolute risk aversion (CARA) preferences and takes the form 
\[ 
    u(x) = \frac{1 - e^{-\alpha x}}{\alpha}
\] 
The coefficient of risk aversion 
\[
    -\frac{u''(x)}{u'(x)} = \alpha
\]
for this function, where $u'(x)$ and $u''(x)$ are, respectively, the first and second derivatives of the utility function. The choice of utility function would be important for real-world deployment of our model, but is a minor consideration for introducing the model. 

A simple means of pricing an insurance contract is to fix the expected utility as a function of payoffs in two states: loss and no-loss (see, for example, Rees and Wambach (2008)~\cite{reesandwambach}). As stated in Section~\ref{subsec:model-structure-params}, for the insurer in this model, this is 
\begin{equation}
    U(\pi_i,P_i,L_i) = \sum\limits_{i} \pi_i u_i (P_i - L_i) + (1 - \pi_i)u_i(P_i)
\label{eq:insurer-utility}
\end{equation}
where the suffix, $i$, represents the different components of the organizational architecture. While Equation~\ref{eq:insurer-utility} may appear simple, the calculation of $\pi_i$, the probability of a loss and $L_i$, the size of the loss are non-trivial. We now proceed to explain how these may be derived from entity relationship diagrams via security maturity models using the payroll example in Figure~\ref{fig:payroll-erd-with-c-and-s}.  

The security maturity model has two key effects on the insurance assessment: how \emph{likely} and how \emph{large} might losses be given the maturity assessment of the organizational architecture. Assume that the probability of a loss, $\pi_i$ may be represented as 
\begin{equation}
    \label{eq:pi-def}
    \pi_i \sim D(\mu_i, C_i, S_i)
\end{equation}
where D is a distribution function, relating the expected probability of loss L to criticality, sensitivity and the maturity function, $\mu_i$ 
(where $\sim$ denotes `is distributed as'). Losses are given by 
\begin{equation}
\label{eq:loss-function}
    L_i(\mu_i,C_i,S_i) = \lambda_i(\mu_i,\Delta C_i,\Delta S_i) 
\end{equation}
\begin{equation}
\label{eq:delta-defs}
    \Delta C_i = C_i - \bar{C_i}, \quad \Delta S_i = S_i - \bar{S_i}
\end{equation}
where $\lambda_i$ is a function specific to the insurer for estimating economic losses from deviations in criticality ($C_i$) and sensitivity ($S_i$) informed by the maturity function, $\mu_i$. $\bar{C_i}$ and $\bar{S_i}$ represent, respectively, the intended state of criticality and sensitivity. These states will vary across industries, with varying emphases on specific risks.   

The reason for having different $\lambda$ functions for each insurer is different insurers may have different information, experience or appetite for particular risks. For instance, one insurer might be highly exposed to one industry and thus there is negative utility associated with writing further insurance to that sector; for another, it might diversify the existing portfolio and thus have positive utility. 

\begin{figure}[ht]
	\centering
	\includegraphics[width=16cm]{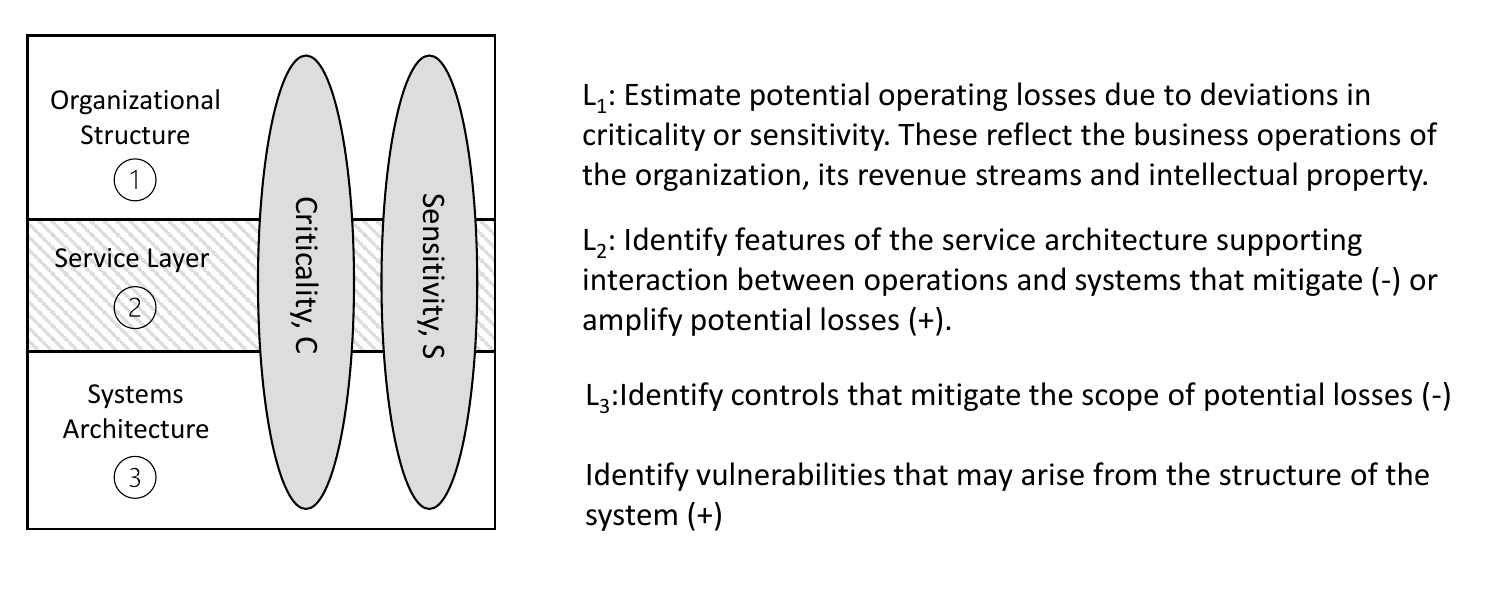}
	\caption{Procedure for estimating potential losses}
	\label{fig:insurance-loss-estimation-cycle}
\end{figure}

It is important to note that $L_i$ is not required to be strictly positive for all layers of the model. The operations of the organization, described by the management layer, determine the scope of possible losses to the business that may result from deviations in criticality and sensitivity as previously described. However, it is possible that additional maturity in terms of practices at the level of the systems layer may reduce the scope of potential losses. Therefore, it is possible that the ultimate components of the premium calculated by the insurer related to the service and systems architecture layers, $P_2$ and $P_3$ respectively, may be negative depending on the assessment of maturity. This is consistent with the practice of insurers offering discounts on baseline policy quotes for meeting certain conditions. However, it may be the case that certain elements of the architecture create greater risks (for example, external connections to a database), and therefore merit charging more premium in that case. Figure~\ref{fig:insurance-loss-estimation-cycle} illustrates how the organizational representation first introduced in Figure~\ref{fig:paper-overview} outlines a possible procedure for estimating the parameters $L_i$ for the different layers of the organizational structure. 

One advantage of the systematic ERD-maturity-utility modelling framework is that once calibrated by an insurer according to its pricing requirements, the framework could provide consistent comparisons between different insurance risks. This allows for its potential use in accumulation modelling, sometimes known as clash exposures, in the insurance industry. The aim of this modelling is to determine the risk of correlation in losses if certain categories of event occur. A well-diversfied insurance portfolio should not contain policies with high levels of clash risk. This is a potential problem for cyber-insurance in particular as some of the usual diversification strategies such as insuring organizations in a wide range of locations, industry groups and annual revenue may be less effective given the interconnected nature of cyber-space and the concentration of computer operating systems. 

The expected-utility-maximization approach adopted here for pricing premia is compatible with other established methodology and models in the literature. The Gordon and Loeb (2002)~\cite{gordon2002economics} Model relates the probability of a security breach and investment in cyber-defences. The Gordon-Loeb security breach functions could be used as the probability distribution for losses in Equation~\ref{eq:pi-def}. This has precedent in the literature, such as in Mazzoccoli and Naldi (2020)~\cite{mazzoccoli2020robustness} and Skeoch (2022)~\cite{skeoch2022expanding}. For the premium pricing model here, one could easily replace investment, $z$ in the Gordon and Loeb model with the maturity function $\mu$ which could be justified by the notion that increased investment in security should bring increased security maturity; otherwise, the investment is not a rational activity. Beyond the Gordon and Loeb Model, Mazzoccoli and Naldi (2022)~\cite{mazzoccoli2022overview} provide a helpful survey of security breach probability models that might aid a corporate decision maker or insurer model the relevant threats. This approach over time could be blended with a distribution bootstrapped from experience cyber-losses but any available datasets currently have significant issues with heterogeneity of losses and limited time coverage rendering meaningful statistical inference difficult. 

The means of estimating the distribution of expected losses for each layer is non-trivial and there are two possible approaches --- construct a distribution \textit{a priori} or fit one from observed data. The latter is what the actuarial department of an insurance company specializes in. For expositional simplicity, we will use the security breach functions (SBF) from the Gordon \& Loeb (2002) model~\cite{gordon2002economics} and specifically, type 1 functions. In the Gordon \& Loeb Model, it is assumed that an information set has an inherent probability of breach or vulnerability, $v$, which is reduced by investment, $z$. The type 1 SBFs take the form 
\begin{equation}
S^I(z,v)=\frac{v}{(\alpha z + 1)^\beta}
\end{equation}
$\alpha$ and $\beta$ are measures of the productivity of information security. For the purposes of populating the utility function, we assume that the probability of a breach follows the Gordon \& Loeb type 1 SBF, but replace investment with the met-objectives parameter, $o_i$, from the maturity model. This is justified by the assumption that to achieve objectives, the insurance-seeking firm must have made an investment in security to implement the practices that allow the objective to be deemed met: 
\begin{equation}
\pi_i = \frac{v(C_i, S_i)}{(\alpha_i o_i +1)^\beta_i}
\end{equation}
The higher the met-objectives parameter, the lower the probability of a breach from the insurer baseline and the lower the premium. $v(C_i, S_i)$ represents the insurer's best estimate of loss probability for the insured component. $\alpha_i$ and $\beta_i$ can be adjusted as required to reflect the insurer's confidence in the efficacy of the objective, or updated over time based on experience. This allows the fundamental form of the analysis and contract to be preserved while allowing for an update on the pricing model based on experience. 

The final component required to be defined is the loss function. To compute the loss function, we assume that the meeting of practices stipulated in the maturity model limits the potential for losses to arise via lowering the potential degradation of criticality and sensitivity. Then 
\begin{equation}
\lambda_i(\mu_i,\Delta C_i, \Delta S_i) = \frac{\Lambda^{(C_i)}\Delta C_i + \Lambda^{(S_i)}\Delta S_i}{(1+p_i)^{\gamma_i}}
\label{eq:onlineretail-utility1}
\end{equation}
$\Lambda$ represents the maximum potential estimated loss to the layer from a degradation in confidentiality or sensitivity, while $\gamma$ is a parameter for the productivity of effectiveness of practices. This might be aligned to the maturity parameter, $m_i$, for example. 

The final step required to populate the utility function is to consider policy limits and coverage. Until now, we have considered the insurance premium purely in monetary terms, which is sufficient for a conceptual overview of the utility function. In reality, the premium is usually decomposed into two components: a premium rate, $r$ (usually expressed in \%) and an amount of coverage (also known as a limit), up to which losses will be reimbursed. The implementation of the limit may vary; in this example, we assume that there is a limit for each layer of the organization for consistency with the model. If we assume that coverage is a function of maturity, then

\begin{equation}
    P_i = r_i \kappa_i m_i
\end{equation}
where $\kappa_i$ represents the policy limit determined by the insurer for the assessed component. 

 For simplicity, let us assume that $u_i(x) = x$ and for this example that utility is linear.\footnote{Optimizing the commonly used economic utility functions is typically algebraically intractable and requires numerical methods.} Taking the above definitions and simplifying Equation~\ref{eq:onlineretail-utility1}, we may now write the utility function
\begin{equation}
U = \sum_i r_i m_i \kappa_i - \frac{v(C_i, S_i)}{(\alpha_i o_i +1)^{\beta_i}}.\frac{\Lambda^{(C_i)}\Delta C_i + \Lambda^{(S_i)}\Delta S_i}{(1+p_i)^{\gamma_i}}
\label{eq:onlineretail-utility-final}
\end{equation}
with the constraint

\begin{equation}
\Lambda^{(C_i)}+\Lambda^{(S_i)} \leq m_i \kappa_i
\end{equation}

Estimating the loss component of the cyber-insurance premium pricing function is less well developed in the literature. This is likely in part because this task is very dependent on the organization in question and the value of its information assets and thus it is hard to model in generality. Woods et al. (2017)~\cite{woods2017mapping} describe the mapping of questions on insurer proposal forms with existing ISO security standards; this suggests that there is some evidence from industry practice that the idea of using security maturity models for pricing insurance contracts has merit. Likewise, Romanosky et al. (2019)~\cite{romanosky2019content}, describe how insurance companies tend to use factors and modifiers for policy pricing based on observations about the insured entity. The use of maturity models to systematically produce such modifiers would be more objective and if standardized across insurers might lead to more efficient risk transfer.  

It is important to note the following: insurance companies have long experience of the magnitude and 
frequency of losses that arise in organizations based on their size, industry sector, and location. Consequently, 
their calculations of premia will start from a baseline determined by these considerations. The contribution of
the methodology proposed here is to provide a framework for calculating the effects of cyber-based risk on the 
frequency and magnitude of losses. This is achieved through a security analysis of the relationship between the operational structure of an organization and its information systems. 


\section{Using the Modelling Framework} \label{sec:case-studies}
\subsection{Case Studies}

This section provides a thought experiment in which we explore how our methodology applies to a range of organizations with differing threat environments and security postures. The analysis is structured to correspond to the modeling approach depicted in Figure~\ref{fig:paper-overview}. First, the operations of the prototypical organization are described from a security perspective. The key elements of security maturity in relation to protecting the criticality and sensitivity of the operations, service, and systems layers of the organization are then discussed. Finally, the potential resultant economic losses from degradation in criticality or sensitivity are considered. In real world examples, two prominent threats are data breaches and ransomware, which compromise sensitivity and criticality respectively although some ransomware attacks also involve data leakage in which case both criticality and sensitivity are compromised.  

In this thought experiment, we give an analysis of three prototypical (fictional) organizations from three different industries: 
\begin{itemize}
    \item Online Retail
    \item Consumer Banking
    \item Pharmaceuticals Manufacture.
\end{itemize}
These industries have a common basic requirement for information security yet the emphasis and implementation will be different in each. In \emph{online retail}, availability of the elements of the organizational structure that process and fulfil sales is paramount. For the \emph{consumer bank}, there is an equal emphasis on confidentiality and integrity; while availability is important, it arguably cannot come at the expense of risking loss of customer funds. Within \emph{pharmaceuticals manufacture}, integrity is of primary importance in the manufacturing process; in research and development, confidentiality is also very important. In addition to the specific details relevant to each, all three industries are customer facing and therefore have external information channels, which might be exploited by an attacker. 

The intention with the following case studies is to show how the model introduced in the previous sections might be applied across a range of potential organizations. This flexibility is important for an insurance company that is likely to insure organizations from different industries across a single line of coverage such as cyber-insurance. We make no claim that the analysis that follows represents the optimal security posture for the different industry groups, which would of course require a thorough analysis of their precise systems  and operational practices. Rather, the following analyses should be considered as stylized examples. The prototypical organizations are constrained to three departments which prioritize, respectively, confidentiality, integrity and availability. As explained in Section~\ref{subsec:criticality-sensitivity}, integrity may be considered as the intersection of criticality and sensitivity. For expositional simplicity, only basic features of ERD notation are used as they provide sufficient detail for demonstrating the use of the model in the case studies. This does not preclude the use of richer notation to describe complex real-world organizations such as those with delicate intellectual property considerations, for example. The organizational departments are \emph{entities}, which sit in the operations layer of the entity relationship diagram. Each department has an associated \emph{entity} in the systems layer, connected by an \emph{entity} in the service layer. The interaction between each \emph{entity} is described by \emph{relationships}. \emph{Attributes} convey descriptions or details of the entities or relationships that are important for the maturity assessment. This allows for potential attack surfaces to be quickly identified with a cursory read of the ERD for each organization.            

In each of the case studies, both a maturity assessment and a computation of premia is required. We provide these in some, though not complete, detail for the case of \emph{online retail} and remark that similar analyses can be established for each of \emph{consumer bank} and \emph{pharmaceuticals 
manufacture}, the details of which are, for brevity, elided. 

\subsection{Online Retail}
\subsubsection{Operations}
\begin{figure}[!htb]
\begin{center}
    \begin{tikzpicture}
    \instantiateerd
    \onlineretailerd
    
    \end{tikzpicture}
    
\end{center}
\caption{Entity relationship diagram for online retailer}
\label{erd:online-retail}
\end{figure}
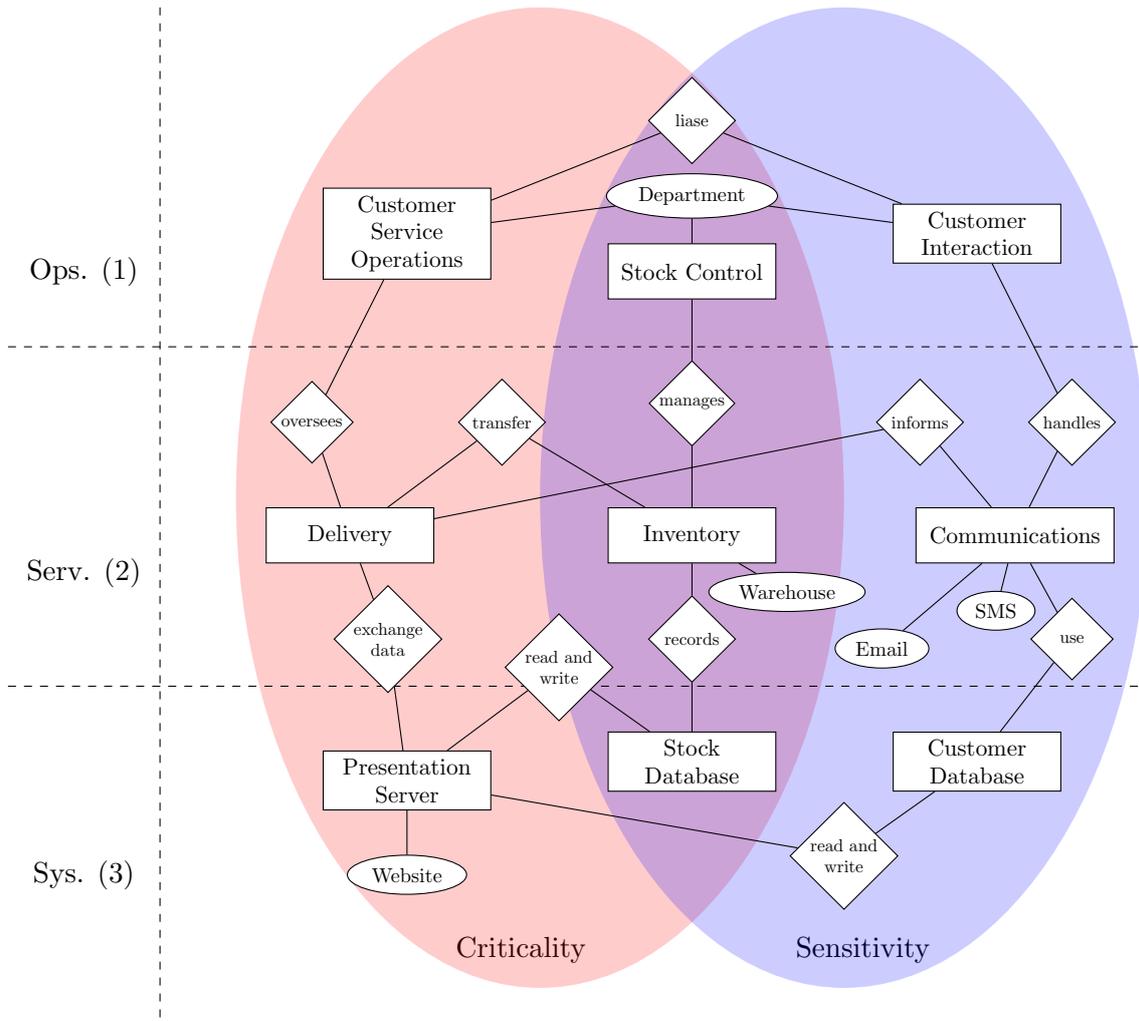

The online retailer in this example uses a website to market and sell goods to customers. The operations of the retailer (Figure~\ref{erd:online-retail}) are represented by the entities \emph{customer service operations}, \emph{stock control}, and \emph{customer interaction}, which respectively require availability, integrity and confidentiality to be protected. For a retailer, the foremost security priority is availability/criticality as this presents the greatest potential source of revenue losses and therefore would be the risk that the retailer would place greatest emphasis on insuring against. Such insurance is known as business interruption. 

Confidentiality is also important for the online retailer given that it handles large quantities of customer data including payment details that may be required to be protected by legislation in the countries in which it operates. This places the online retailer at risk of a regulatory fine for data breaches and also loss of reputation of customer confidence. However, this risk is secondary in terms of immediate potential direct losses relative to availability. The customer interaction department is responsible for dealing with customers. There is a distinction to be made however, between communications such as marketing and automatic notifications.

Integrity is the lowest security priority for the online retailer. The prototypical retailer has a stock control department which is responsible for recording the stock inventory for the retailer. In an optimally run retailer, this would be robust and in real time, but this provides an illustration of the concept of maturity in this context. In a retailer with low maturity of stock control, updates of inventory might be manual and thus there is potential risk of an out-of-stock item being sold to a customer. By contrast, a high volume, established online retailer will likely have sophisticated inventory management technologies that aim to prevent such a scenario from occurring. However, loss of integrity is an isolated business risk in this case and would therefore likely receive relatively little priority in insurance terms.  

\subsubsection{Maturity assessment} \label{subsec:online-retailer-maturity-assessment}

As stated in Section~\ref{subsec:model-structure-params}, a requirement of our model is that the analysis of the system's security posture provided by the maturity model must support the instantiating of the required utility function for the insurance contract. The analysis must deliver parameters representing the maturity of protecting the security parameters criticality and sensitivity, which as previously described are delivered via practices organized into domains and objectives. For the online retailer, the most important security parameter is criticality and specifically the level of safeguards or redundancy in the system that may keep the operations available. The second most important output is an assessment of the protection of confidentiality and the scope for loss of data. The cost of such potential breaches should be treated separately within the utility function rather than the maturity model. 

For an online retailer, one might start the maturity assessment by identifying domains as a baseline for calculating the extent to which the security posture is achieved: 

\begin{enumerate}
    \item Cloud usage
    \item Incident response
    \item Access Control
    \item Maintenance
    \item System and Communications Protection
    \item System and Information Integrity.
\end{enumerate}
These are intended to provide the necessary set of domains to instantiate the utility function of the insurer rather than provide a complete description of the online retailer's security maturity. For this example, we will not stipulate individual practices within the domains as these are specific to the precise implementation of the system and controls. To move from domains to maturity, we will consider a number of sample objectives, which may also be regarded as insurability criteria. The extent to which these objectives are met by an insurance client in turn determine the maturity and thus the discount or premium relative to the insurer's standard baseline. Table~\ref{table:online-retail-objectives} states a few sample objectives, which are intended to serve as examplars rather than comprising an exhaustive set of objectives. In a full insurance assessment, a substantive matrix of objectives and assessment criteria would be required. 

Within our model, the maturity function is calculated for each layer --- Management/Operations, Services, and Systems --- of the organization. One resultant consideration for the insurer is how to measure the required parameters of the maturity function for each layer and thus calculate them. For the operations/management layer, this is most likely achieved via asking structured questions of the organization or inspecting existing policies should the insurance buyer be willing to share them. The service layer would require a combination of the approach for the operations/management layer and a systematic evaluation of controls. For the systems layer, the controls can likely be verified more rigorously via, for example, perimeter scanning of defences. The assessment of the operations/management layer is the most subjective and therefore most uncertain and that of the systems layer the most objective and therefore most certain. This uncertainty will be reflected in the formulation of the utility function in Section~\ref{subsubsec:retail-premia}. 

\begin{table}[hb]
    \begin{center}
    \begin{tabular}{|r|p{6cm}|p{6cm}|}
    \hline
    & \textbf{Criticality} & \textbf{Sensitivity} \\ \hline
    \textbf{Management} &(C1) Incident Response plan in place and tested & (S1) Information controls deployed \\
    && (S2) Staff training on handling sensitive information enforced and documented \\
    \hline
    \textbf{Service} &(C2) Customer facing services must have minimal downtime & \\ \hline
    \textbf{Systems} &(C3) Systems should not have vulnerable points of failure &(S3) Databases must be secured against unpermitted modification \\ \hline
    \end{tabular}
    \end{center}
\vspace{1mm}
\caption{Objectives for online retailer maturity}
\label{table:online-retail-objectives}
\end{table}

Table~\ref{table:online-retailer-domain-objectives-matrix} provides domain importance ($d_i$) parameters for the objectives listed in Table~\ref{table:online-retail-objectives}.
\begin{table}[ht]
\begin{center}
    
	\begin{filecontents*}{online-retailer-domain-objective-matrix.csv}
Domain,C1,C2,C3,S1,S2,S3
Cloud Usage,1,1,1,1,0,1
Awareness and Training,1,0,0,0,1,0
Incident Response,1,1,0,0,0,1
Access Control,0,0,0,1,0,1
Maintenance,0,1,0,0,0,0
System and Communications Protection,0,1,1,1,0,1
System and Information Integrity,0,1,1,0,0,1
\end{filecontents*}
\pgfplotstabletypeset[
	col sep = comma,
	columns = {Domain,C1,C2,C3,S1,S2,S3},
	columns/Domain/.style={verb string type, column name={Domain}, column type={|l|}},
	columns/C1/.style={postproc cell content/.code={
	    \ifnum##1=1 \pgfkeysalso{@cell content=\cellcolor{lightgray}{##1}}\fi}},
	columns/C2/.style={postproc cell content/.code={
	    \ifnum##1=1 \pgfkeysalso{@cell content=\cellcolor{lightgray}{##1}}\fi}},
	columns/C3/.style={postproc cell content/.code={
	    \ifnum##1=1 \pgfkeysalso{@cell content=\cellcolor{lightgray}{##1}}\fi}, column type={c|}},
	columns/S1/.style={postproc cell content/.code={
	    \ifnum##1=1 \pgfkeysalso{@cell content=\cellcolor{lightgray}{##1}}\fi}},
     columns/S2/.style={postproc cell content/.code={
	    \ifnum##1=1 \pgfkeysalso{@cell content=\cellcolor{lightgray}{##1}}\fi}},
	columns/S3/.style={column type ={c|},postproc cell content/.code={
	    \ifnum##1=1 \pgfkeysalso{@cell content=\cellcolor{lightgray}{##1}}\fi}},
	every head row/.style={before row=\hline,after row=\hline},
	every last row/.style={after row=\hline}
	]{online-retailer-domain-objective-matrix.csv}
    \vspace{1mm}
	\caption{$d_i$ parameters for online retailer maturity}
	\label{table:online-retailer-domain-objectives-matrix}
	\end{center}
\end{table}
As described in Section~\ref{subsec:model-structure-params}, it is important to understand how the analysis presented above depends upon the specific interdependencies in the system. In this 
case, we can see this by considering Table~\ref{table:online-retail-objectives}. 

In each of the components of Table~\ref{table:online-retail-objectives}, the $C$ and $S$ values are determined not only by the entities directly concerned but also by the relationships of these
entities to other entities as specified by the ERD for the online retailer's system (see Figure~\ref{erd:online-retail}). For example, the criticality assessment of the `Presentation 
Server', in the System Layer, is dependent on the criticality assessment of  `Delivery' in the 
Service Layer and, also in the Service Layer, on the sensitivity assessment of the `Customer Database', and the integrity assessment of the `Stock Database'; and so on. 

\subsubsection{Economic losses}

The online retailer will primarily require coverage for losses from business interruption. In an insurance policy this might be specified as coverage up to a certain number of business days with an initial deductible. Coverage may also be required for potential legal and investigative costs associated with a data breach and loss of customer data. In order for the insurer to compute a premium for the policy, as per Equation~\ref{eq:insurer-utility}, there are two variables needed for each layer, $i$, of the ERD representation for the online retailer (Figure~\ref{erd:online-retail}): $\pi_i$, the probability of a loss, and $L_i$, the expected magnitude of a loss. As emphasized in Section~\ref{subsec:cyber-insurance-policies}, the maturity model is expected to deliver the potential deviation from the baseline assumptions of the insurance company for the firm in question.

\subsubsection{Computing Premia} 
\label{subsubsec:retail-premia}

As discussed in Sections \ref{subsec:model-structure-params} and \ref{subsec:cyber-insurance-policies}, the utility equation for the insurer takes the form

\begin{equation}
U = \sum_i r_i m_i \kappa_i - \frac{v(C_i, S_i)}{(\alpha_i o_i +1)^{\beta_i}}.\frac{\Lambda^{(C_i)}\Delta C_i + \Lambda^{(S_i)}\Delta S_i}{(1+p_i)^{\gamma_i}}
\label{eq:onlineretail-utility-final-repeat}
\end{equation}

To populate the utility function parameters, the approach discussed here will require the maturity, $m_i$ of components; the extent to which practices, $p_i$, in the maturity model are met; and the extent to which objectives, $o_i$, specified by the insurer in the maturity model are met. For simplicity in this example, we assume that $\bar{C_i} = \bar{S_i} = 1$ and that both $C_i$ and $S_i$ take values across $[0,1]$. This provides a simple means of combining these parameters with a potential maximum loss or policy limit.  

Table \ref{table:online-retail-u-params} illustrates the parameters that might be used to populate Equation~\ref{eq:onlineretail-utility-final-repeat}. The following assumptions are made:
\begin{itemize}
    \item Simulations would be run to estimate deviations in criticality and sensitivity under various scenarios. A starting point would be Lloyd's of London realistic disaster scenarios (RDS) for cyber incidents. 
    \item The values for maturity, practices, and objectives are purely indicative in this example. For simplicity, in this example, the parameters representing maturity, practices met, and objectives met are set to the same values for each assessed component. As discussed in Section~\ref{subsec:online-retailer-maturity-assessment}, the objectivity of assessment and accordingly ability to deem practices met increases from the operations layer, through the services layer, to the systems layer. The illustrative values for $m_1 < m_2 < m_3$ are deliberately chosen to reflect this. 
    \item The maximum potential loss per layer is purely illustrative and the policy limit for each layer in the example is set to this. 
    \item The loss probabilities are purely illustrative; an insurer would either estimate these using actuarial methods or based on real loss experience. 
    \item The productivity parameters are set to 1 for simplicity. The online retailer operates a fairly open system that prioritizes criticality and thus the efficacy of controls may be reduced relative to a more closed system.
    \item The premium rate would be solved for using standard numerical methods of choice. The focus of this discussion is to illustrate how to produce an assessment, not to estimate specific premia.  
\end{itemize}
The preceding analysis is intended only to provide an indication of how a utility function might be constructed. However, the broad form should be applicable across the case studies. The equation could potentially be solved analytically using readily available software; for more complex formulations, Monte Carlo simulations of losses would be an acceptable strategy. 
 \begin{table}[!htb]
\begin{center}
    \begin{tabular}{|l|l||c|c|c|}
        \hline
        Parameter & Description & Cust. Serv. Ops. (1) & Inventory (2) & Cust. Data (3) \\
        \hline
        $m_i$ & Maturity & 0.5 & 0.6 & 0.7 \\
        $p_i$ & Practices met & 0.5 & 0.6 & 0.7 \\
        $o_i$ & Objectives met & 0.5 & 0.6& 0.7 \\
        \hline
        $\Lambda_i$&Maximum potential loss per layer& \$100k&\$200k&\$300k\\
        $\kappa_i$& Policy limit & \$100k&\$200k&\$300k \\
        \hline
        $v(C_i, S_i)$&Loss probability&0.05&0.02&0.01\\
        $\alpha_i,\beta_i,\gamma_i$&Productivity parameters&1&1&1\\
        \hline
        $\Delta C_i, \Delta S_i$&Potential degradation in C and S&\multicolumn{3}{l|}{Derived from simulated scenarios, e.g. Lloyds RDS}\\
        \hline
        $r_i$ & Premium rate & \multicolumn{3}{l|}{Outputs of chosen solution method of utility function}\\
        \hline

    \end{tabular}
\end{center}
\caption{Summary of utility function parameters for online retail example}
\label{table:online-retail-u-params}
  \end{table}
 
\subsection{Consumer Bank}
We represent the prototypical consumer bank via the entities \emph{records}, \emph{payments} and \emph{customer interaction}, for which respectively availability, integrity and confidentiality are required to be protected. The basic operation of the bank is as follows: customers register transactions either withdrawing or depositing funds. These transactions are mandated via interfaces, such as banking cards or online banking facilities. The transactions are processed by the payments department and, if validated, records are then updated. 

In terms of security objectives, integrity of the payments system is marginally most important as manipulation of this entity places the bank at greatest financial risk. Confidentiality of customer data is also of high priority, in particular appropriate protocols need to be in place to ensure customer accounts are resistant to misuse. This is typically achieved through the use of password mechanisms and other means of verifications that have become more sophisticated over time. Finally, availability of the records system need to be protected. While this is something a bank would strive to achieve, it is not uncommon of for banks to implement system downtime for maintenance while allowing routine payment processing to continue to take place. Further, in many banking systems, there is a lag between a payment being taken and its being recorded; for credit card transactions, this can be several business days. 

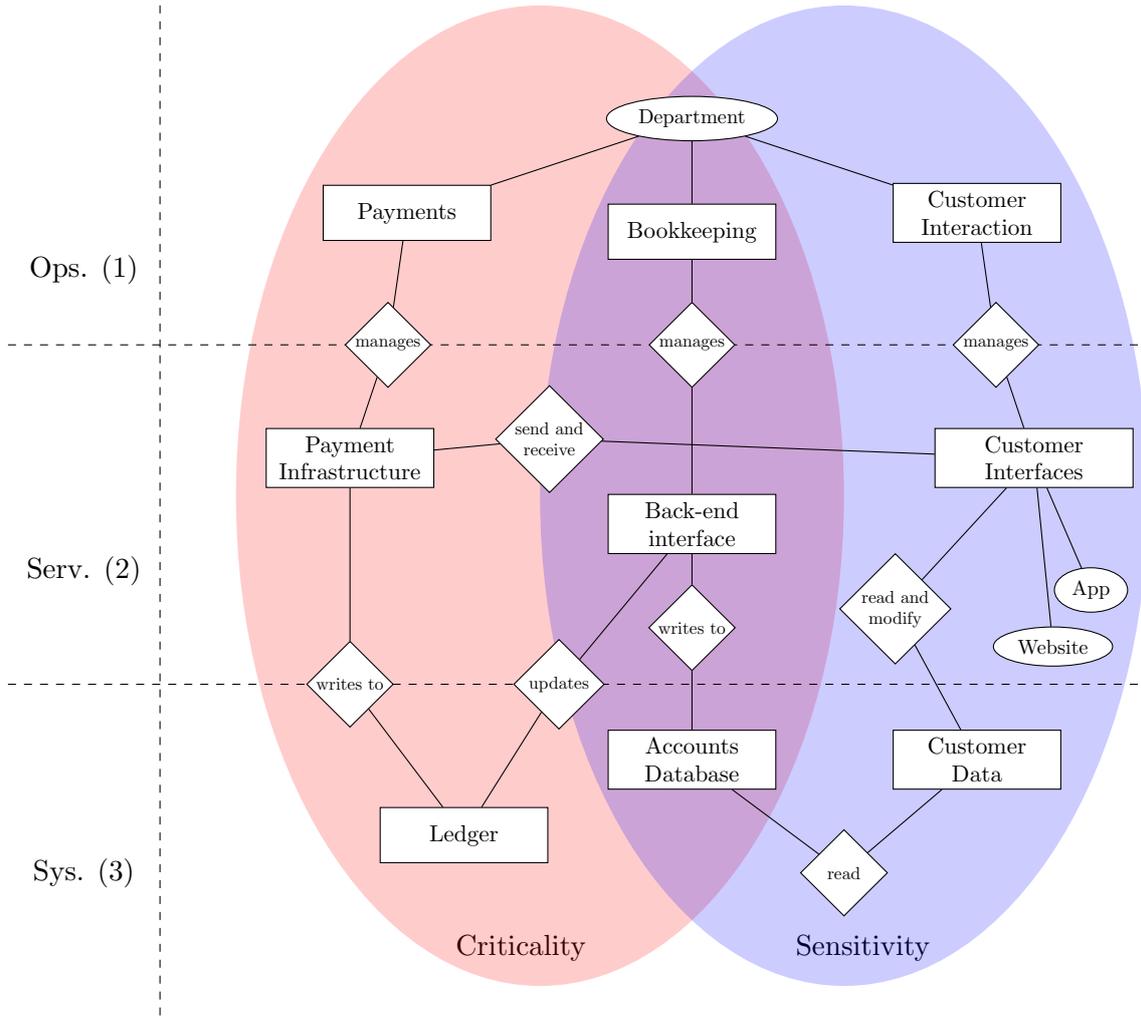
\begin{figure}[ht]
\begin{center}
    \begin{tikzpicture}
        
    \instantiateerd
    \bankingerd

    \erdshading
    
    \end{tikzpicture}
    
\end{center}
\caption{Entity relationship diagram for consumer bank}
\label{erd:bank}
\end{figure}

\subsubsection{Maturity assessment}

We proceed to outline the maturity assessment process for the consumer bank in a similar fashion to Section~\ref{subsec:online-retailer-maturity-assessment} for the online retailer, though with a lesser level of detail. The issue of banking security is complex and well-studied, and in this relatively brief discussion it is not possible to conclusively cover the issue. Chapter 12 of Anderson (2020)~\cite{anderson2020security} provides a lively discussion of both the history of security models used in banking and contemporary methodology. The priority of the assessment is to deliver how susceptible the banking system is to integrity attacks. A formal model applicable to the security policy of a bank was proposed by Clark and Wilson (1987)~\cite{clark1987comparison}. In brief, the model describes how an unconstrained data item (UDI) may be transformed into a constrained data item (CDI). The integrity of a CDI must be preserved, which is validated through integrity verfication procedures (IVPs). Modification of CDIs is effected via a transformation procedure (TP), which under some circumstances may transform a UDI into a CDI. 

Handling and mitigating losses is routine activity for a consumer bank; the role of cyber-insurance is to cover losses outwith the ordinary course of business that are generated either by a failure in technology or circumvention of controls. Effectively, the relevant parameters the maturity model for the consumer bank ought to deliver are:

\begin{itemize}
    \item How likely is a catastrophic event?
    \item What might it cost?
\end{itemize}

As noted in the aforementioned discussion by Anderson, malicious insider risk is one of the foremost potential loss generators for a consumer bank. One area of insurance coverage that is relevant for a bank is technology errors and omissions coverage. This would deliver compensation for damages caused by coding or system configuration errors. Such an assessment would need to be kept highly confidential as it otherwise might provide a `how-to manual' for potential attackers. Referring to Figure~\ref{erd:bank}, the \emph{Customer Interfaces} are a potential point of weakness. If sensitivity of these is not properly protected, a malicious insider might be able to impersonate a customer to misdirect funds. In the event the malicious insider has sufficient access to credentials, it may be hard for the bank to separate fraudulent from genuine activity. This suggests that the maturity model domain `Personnel Security' would be of high priority for all three \emph{Department} entities within the bank. 

Within the systems layer, preservation of the \emph{Accounts Database} is of the utmost importance. A potential avenue of attack would be for an employee in \emph{Bookkeeping} to manipulate the back-end interface to divert funds from accounts in modest quantities (`skimming'). In the service layer, criticality of \emph{Payment Infrastructure} must be preserved. As seen in Figure~\ref{erd:bank}, \emph{Payment Infrastructure} has the relationships \emph{writes to} with \emph{Ledger}, \emph{updates} with \emph{Back-end Interface}, \and \emph{send and receive} with \emph{Customer Interfaces}. It is clear from these relationships that a failure of \emph{Payment Infrastructure} would cause significant issues across the organisation. Therefore, it would need to be given high priority in the maturity assessment of the bank.  

\subsubsection{Economic losses}

The largest risk the consumer bank faces on a regular basis is cyber-enabled fraudulent activity. This may be direct or indirect --- using the bank's customers as a conduit. The elements of the bank with significant exposure to criticality are likely to be closed systems; interbank systems are usually not directly exposed to the internet and are will be rigidly firewalled. Given the importance of availability of such systems, banks will usually invest in infrastructure such as uninterrupted power supplies, multiple data centres and disaster recovery sites. The major economic losses a consumer bank might seek coverage against would be disaster recovery beyond that already budgeted for in its operating model; fraud in excess of a certain threshold; recovery from a significant incident causing a material loss of revenue.  

\subsubsection{Computing Premia} \label{subsubsec:banking-premia}

This analysis follows Section~\ref{subsubsec:retail-premia}, but rather than providing a full utility function, we discuss a few relevant considerations an insurer may wish to consider. Recall that the form of the insurer utility function is
\begin{equation}
    U(\pi_i,P_i,L_i) = \sum\limits_{i} \pi_i u_i(P_i - L_i) + (1 - \pi_i)u_i(P_i)
\label{eq:insurer-utility2}
\end{equation}
We now briefly sketch some considerations for populating $\pi_i$ and $L_i$.
\begin{table}[htb]
\begin{center}
    \begin{tabular}{|l|p{4.5cm}|p{4.5cm}|p{4cm}|}
    \hline
    Layer & $\pi_i$ & $\Delta C_i$ & $\Delta S_i$ \\
    \hline
    Ops/Mgmt & Incidence of management failures & Failure of Payment Infrastructure & Customer redress \\
    \hline
    Services & Incidence of insured losses due to fraud & Extortion or ransomware & Strength of controls in customer interfaces \\
    \hline
    Systems & Incidence of system failures & Ledger controls & Customer data controls \\
    \hline 
    \end{tabular}
\end{center}
\caption{Utility function parameters for the consumer bank}
\label{table:utility-fn-params-consumer-bank}
\end{table}
The methodology used for parametrizing the online retailer maturity is arguably too general for the case of the consumer bank. It should be remembered that in many jurisdictions, banks are subject to rigorous stress tests and regulation. In the United Kingdom, for example, there already exists a protocol for testing the maturity of cyber-security for banks, CBEST~\cite{CBEST}. Stress test parameters could be used to populate the utility function; these have the advantage of being measured both in the spaces of probability and losses. One potential insurance strategy might be for the insurer to offer coverage against losses of a probability below a specified threshold (for example, 1\%). An excess-of-loss policy, covering losses $L_{min} \leq L \leq L_{max}$ might be the best starting point for constructing the utility function. Based on the maturity assessment, the other required parameters such as maximum available coverage could then be populated and the premium rate calculated. 

\subsection{Pharmaceuticals}

\subsubsection{Operations}

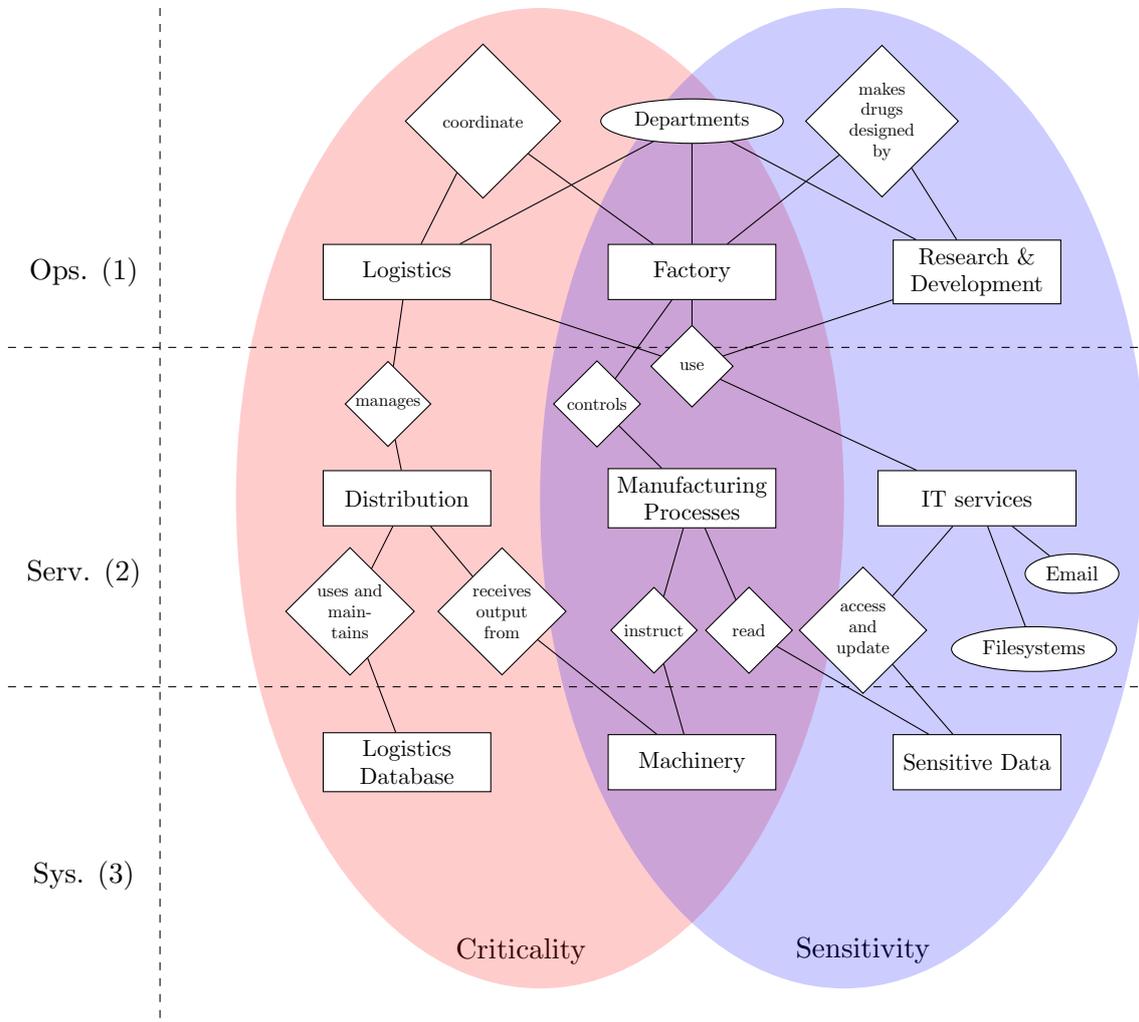
\begin{figure}[htb]
\begin{center}
    \begin{tikzpicture}
        
    \instantiateerd
    \pharmaerd

    \erdshading
    
    \end{tikzpicture}
    
\end{center}
\caption{Entity relationship diagram for pharmaceutical manufacturing}
\label{erd:pharma}
\end{figure}

We represent the operations departments of the prototypical pharmaceutical manufacturer via the entities \emph{logistics}, \emph{factory} and \emph{research \& development}, for which respectively availability, integrity and confidentiality are required to be protected. The foremost priority for the pharmaceutical manufacturer is the integrity of the factory and manufacturing processes as compromise of this potentially places lives at risk. The research and development department will hold sensitive information, both commercial and personal, about potential drug candidates and clinical trials, which must be protected. The availability of the logistics entity is the lowest priority for the pharmaceutical manufacturer compared with the risks of compromise of integrity or confidentiality. The ERD for the pharmaceutical manufacturer is presented in Figure~\ref{erd:pharma}.

\subsubsection{Maturity assessment}

A relevant security model for the pharmaceutical manufacturer is due to Biba (1975)~\cite{biba1975integrity}. The Biba model is essentially a read up, write down model. By means of illustration, consider a drug, which has a specific chemical structure. The goal of pharmaceutical manufacturing is to reliably produce that chemical structure from a range of chemicals according to a specified synthetic procedure. It is evident that the elements of the process responsible for manufacture need to be able to read the synthetic procedure, but there is no reason why such elements should be able to modify the procedure.  

Relative to the other two organizations we have analysed, the pharmaceutical manufacturer operates a relatively closed system. Accordingly, an established `off-the-shelf' model of maturity such as the CMMC would arguably be sufficient for assessing the pharmaceutical manufacturer from an insurance perspective. Particular attention should be directed at the controls in place to protect the sensitivity of the \emph{sensitive data} database. There is also an integrity objective in terms of the interaction between the \emph{manufacturing processes} and the database. Many of the risks faced by the pharmaceutical manufacturer will be covered by traditional insurance policies and therefore the scope of the maturity assessment for the cyber-insurance policy will be very specific. An example is the criticality of the systems and service infrastructure supporting the \emph{logistics} function. If there is disruption in this area as a result of a cyber-incident, it may cost the company revenue and inflict reputational damage. 

\subsubsection{Economic losses}

The pharmaceutical manufacturer has the most closed information systems of the three prototypical organizations considered and also has the greatest potential to put in place rigid controls without meaningfully impeding business operations. From an insurance perspective, it is unlikely that the pharmaceutical manufacturer will experience regular losses that it cannot contain itself and the company itself would most likely want to insurer against rare, catastrophic events with potential to seriously damage its business. 

\subsubsection{Computing Premia} 
\label{subsubsec:pharma-premia}

The computation of premia would follow a similar analysis to Section~\ref{subsubsec:retail-premia} for the online retailer. We refrain from giving a parametrized utility function for the pharmaceutical manufacturer as the procedure and process have been thoroughly demonstrated in the other case studies. The utility function for the pharmaceutical company would involve instantiating the probability of losses for each layer, $\pi_i$, and the expected magnitude of losses, $L_i$. Pricing of the pharmaceutical policy would likely be most efficient using a deductible --- an amount of losses the company itself bears before the insurance company begins to reimburse losses. This would result in a lower upfront premium, which would be rational for a pharmaceutical company which is the most able to protect itself from threats of the three prototypical organizations we have considered.  

\subsection{Reflecting on the Case Studies} 
\label{subsec:reflecting}
\begin{itemize}
\item Our chosen case studies illustrate the applicability of our proposed methodology across 
a variety of industries, with varying risk profiles 
and security postures. 
\item They also illustrate how the varying risk profiles and security postures give rise to premia 
that differ from the baseline of sector profile. 
\item Finally, the parametrization of maturity illustrates how 
domains and practices within a sector can give rise to sector-specific 
accumulation risk. For example, it may be the case that a specific system is widely used in an industry for access control. Should an authentication bypass vulnerability arise, then multiple insured firms in that industry could potentially be affected. This is commonly termed `supply chain risk'. 
\end{itemize}

\section{Discussion}
\label{sec:discussion}
\subsection{Real world considerations for using the modelling framework}
In Section~\ref{sec:case-studies}, we explained how to deploy the model for prototypical organizations. Here, we outline some considerations for a real-world insurance assessment. Implementing the model for an insurance assessment of an organization requires the following steps:
\begin{itemize}
    \item Construct the Entity Relationship Diagram
    \item Choose the maturity model and parameters
    \item Identify and assemble relevant data to perform the maturity assessment and compute required inputs (probabilities and losses) for the utility function 
    \item Estimate the utility function. 
\end{itemize}

Romanosky et al. (2019)~\cite{romanosky2019content} and Woods et al. (2021)~\cite{woods2021county} explain some of the procedures and processes undertaken by insurers in pricing policies. Briefly, most insurance companies have some form of baseline premium relative to a company's size and risk, which is then adjusted for various stated factors such as policy limit, controls in place, prior claims history, and so on. The company may use a questionnaire to gather information that can be used to determine which factors to apply. The actuarial department of an insurance company is responsible for determining the required probability distributions to be used in policy pricing. These will usually be derived from prior claims experienced, but also may reference appropriate external parameters and/or forecasts. Ultimately, the modelling framework proposed in this paper is intended to help assist this process, not replace it, and to help address some of the difficulties in mapping data of past cyber-insurance losses to potential future losses. These include issues such as technological evolution and the dynamic nature of cyber-threats; for example, the emergence of ransomware over the past few years, which prior loss data could not have predicted.

The first step in the insurance assessment is to construct the Entity Relationship Diagram per Section~\ref{subsec:ERDs}, containing only the information that is necessary to guide the maturity assessment and insurance pricing. Categorizing risks in terms of criticality and sensitivity allows for the security posture of the organization, its layers, services, and systems to be quickly interpreted. The choice of maturity model should be such that it allows for meaningful comparison between similar organizations for the purposes of policy pricing. It can also be deployed to be aligned with the factors in existing insurance assessments, but with greater depth, structure, and specificity than some questionnaires available in the public domain. Structuring the assessment in this way also allows for any required data to be clearly identified. Finally, the utility function should be framed so that the probability of expected losses and their expected magnitude can be computed based on the maturity model outputs.  

It is worth reiterating that the primary determinants of potential losses for an organization are the disruption to its revenue generating activities or costs incurred by a cyber-incident rather than a generic estimation of `cyber-damages'. Our modelling framework is thoroughly grounded in the structure of an organization and therefore ensures that the treatment of cyber-losses is proportional and appropriate relative to the organization under consideration. At present, there appears to be disagreement between \textit{ex ante} and \textit{ex post} estimates of cyber-losses. There is a good discussion in~\cite{woods2021county} on this topic, while Eling and Wirfs (2019)~\cite{eling2019actual} provide some distributions calculated on experienced operational cyber-losses. Tatar et al. (2020)~\cite{soa-cyber} have developed a model for the Society of Actuaries that can be deployed in an insurance context to describe the specific attack vectors of an organization's systems that would potentially combined well with the model in this paper. One factor worth considering in relation to \textit{ex ante} losses is the extent of employee training and engagement with organizational security policy. This is covered under the `awareness and training' domain of the CMMC maturity model, and belongs to a set of security considerations commonly termed `human factors'. As malicious emails are a common means of malware delivery by threat actors, the extent to which human factors form a risk of insurance losses arising should be given due consideration when using the modelling framework. 

We would stress that there is no one correct way to deploy the modelling framework and that it has been developed with the aim of being flexible and adaptable to different use cases while providing some consistency of overall approach.

\subsubsection{Compatibility existing cyber-security standards}
Our model is compatible with existing cyber-security standards, which as discussed in Section~\ref{sec:sec-mat-models} form the basis of many maturity models. However, these standards are not perfect, as discussed in Chapter 28 of Anderson (2020)~\cite{anderson2020security}. Anderson contends that standards such as ISO27001 for security management are deficient in engendering improved security outcomes. The reason for this is that firms are audited and supply the auditors with information regarding the controls. This usually relies on the principle of `good faith', but a misunderstanding or lack of knowledge about the efficacy of controls can have catastrophic consequences. As Anderson points out, many firms with large data breaches have been ISO27001 certified, yet the breach still occurs. There are third party firms, such as Bitsight, who offer scanning services, which are popular in the cyber-insurance industry. Yet the security score they provide is only applicable to the systems layer of the model we have proposed, and standards if self-certified by management cover only the operations layer. The model we have suggested provides an integrated approach across the three layers of an organization: operations/management, service, and systems. The model allows for an insurance company to identify potential risks and price the premium appropriately. A company might be motivated to consider the benefits of investment in security to reduce its insurance premium following a maturity assessment. 

\newpage 

\subsection{Weaknesses}

\subsubsection{Developing the Entity Relationship Diagrams}
The Entity Relationship Diagrams for an organization could quickly become complex and careful consideration must be given in a real-world deployment as to how much information to include. A requirement for the model to be realistic is that sufficient data be available to populate the maturity assessment and utility function. We have also negated the risks of asymmetric information and/or moral hazard in our model. These are important considerations for pricing of insurance. 

\subsubsection{Subjectivity in Choice of Maturity Model}
The choice of maturity model determines the pricing outputs of our model. It is possible that a particular maturity model could be chosen to attempt to move pricing in a particular desired direction by an insurer, or a company might use a maturity model that deliberately covers suspected weaknesses in its own controls. 

\subsection{Extensions and refinements}

\subsubsection{Cloud usage}
In the systems assessment, one potential avenue of investigation is the interaction between cloud systems and insurance. The extent of cloud usage is potentially important as some of the potential criticality issues that insurance might be sought to mitigate may be covered by the service level agreement (SLA) with the cloud provider. In the event that the online retailer is fully reliant on cloud-based infrastructure for its systems layer, then a specialized form of insurance cover may also be available.  (Table~\ref{table:cloud-cyber-coverage}). A cloud provider is uniquely positioned to assess the maturity of its customer base, subject to the relevant consents and establishes a synergy with the cyber-insurer.

\begin{table}[htb]
\begin{center}
    \begin{tabular}{|l|l|c|}
    \hline
    Cloud Provider & Insurer & Reference \\
    \hline
    Google & Allianz/Munich Re & \cite{munichrepresser} \\
    Microsoft & At-Bay & \cite{microsoftpresser} \\
    AWS & Cowbell/Swiss Re & \cite{awspresser} \\
    \hline
    \end{tabular}
\vspace{1mm}
\caption{Cloud provider and cyber-insurer collaborations}
\label{table:cloud-cyber-coverage}
\end{center}
\end{table}

\subsubsection{Extending the systems modelling}

A potential extension of the work would be to integrate the model based on ERDs, maturity models, and utility with a strongly compositional approach to systems modelling (again, e.g., among many, \cite{Birtwistle1979,Milner2009,CMP2012,caulfield2015improving}). Such an extension would support the scaling of our approach to larger, more complex systems. 

\subsubsection{ERDs as an insurance template}

There is potential for an insurance company to use ERDs as a template for an insurance assessment along the lines of what has been described in this paper. The nuances of ERD notation could be used to depict what perils an insurance company is willing to cover and which they are not. The approach to constructing such a template is not trivial and is likely to be a considerable research and practical endeavour in its own right. 

\section{Acknowledgements}

This work was supported partly by the UK Engineering and Physical Sciences Research Council grant for Doctoral Training EP/R513143/1 and partly by the UK Engineering and Physical Sciences Research Council grant EP/R006865/1.   

\clearpage

\printbibliography

\clearpage

\begin{appendix}
   
    \section{Entity Relationship Diagram Notation and Symbols}
    \label{appendix:erd-notation}
    \begin{figure}[!htb]
	   \centering
	   \includegraphics[width=16cm]{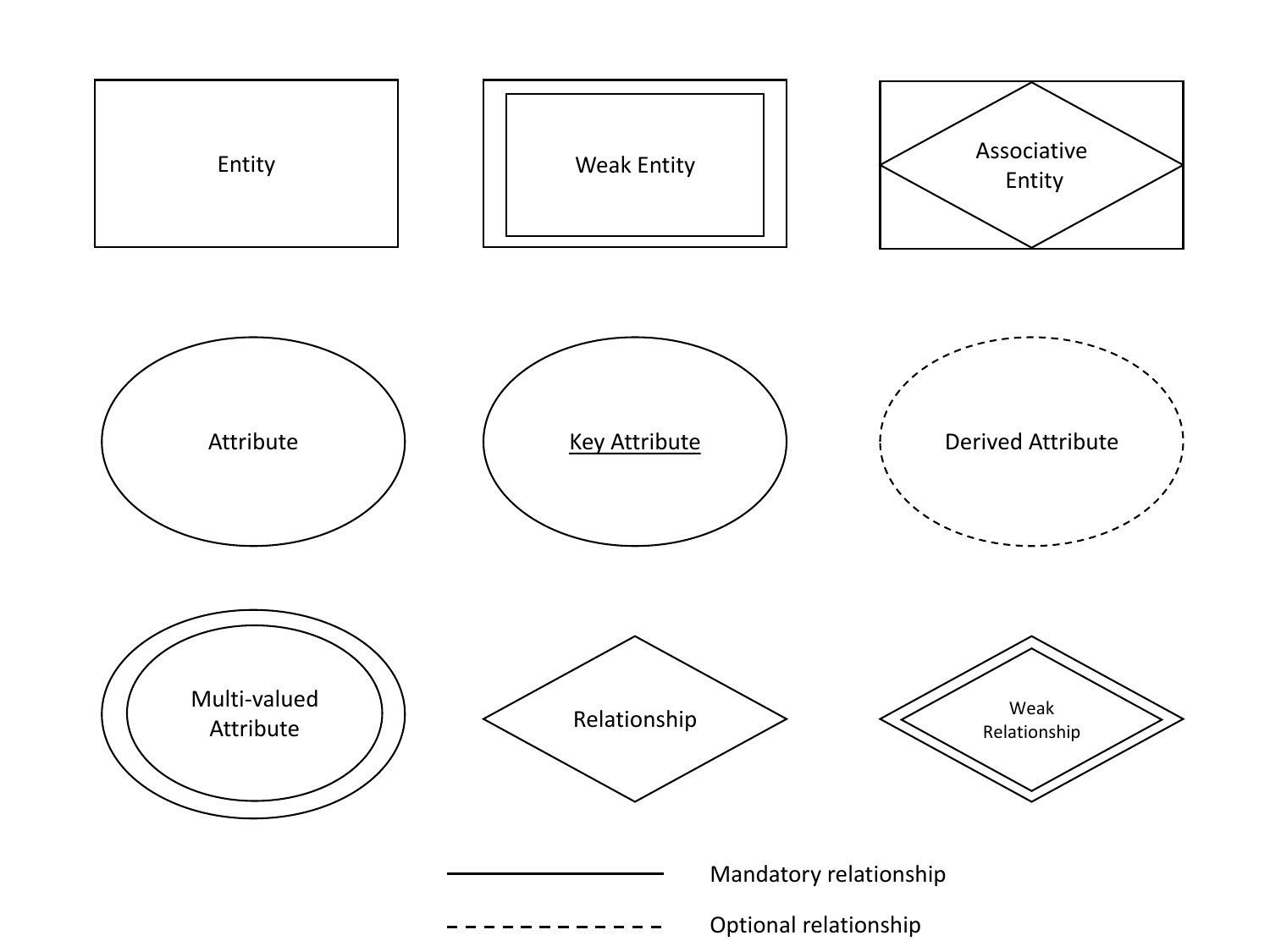}
	   \caption{Chen notation}
	   \label{fig:chen-notation}
    \end{figure}
    
    \begin{figure}[!htb]
	   \centering
	   \includegraphics[width=16cm]{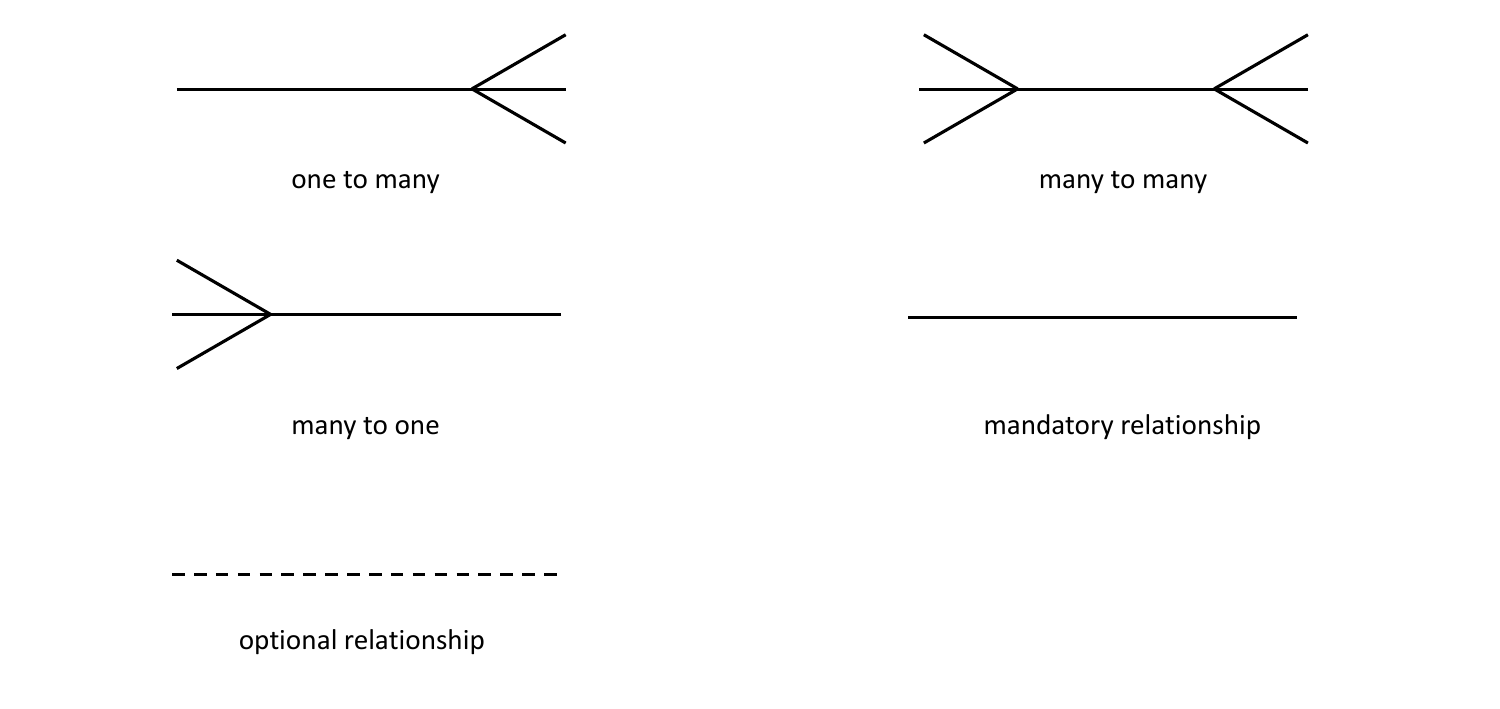}
	   \caption{Barker notation}
	   \label{fig:barker-notation}
    \end{figure}

     \begin{figure}[!htb]
	   \centering
	   \includegraphics[width=16cm]{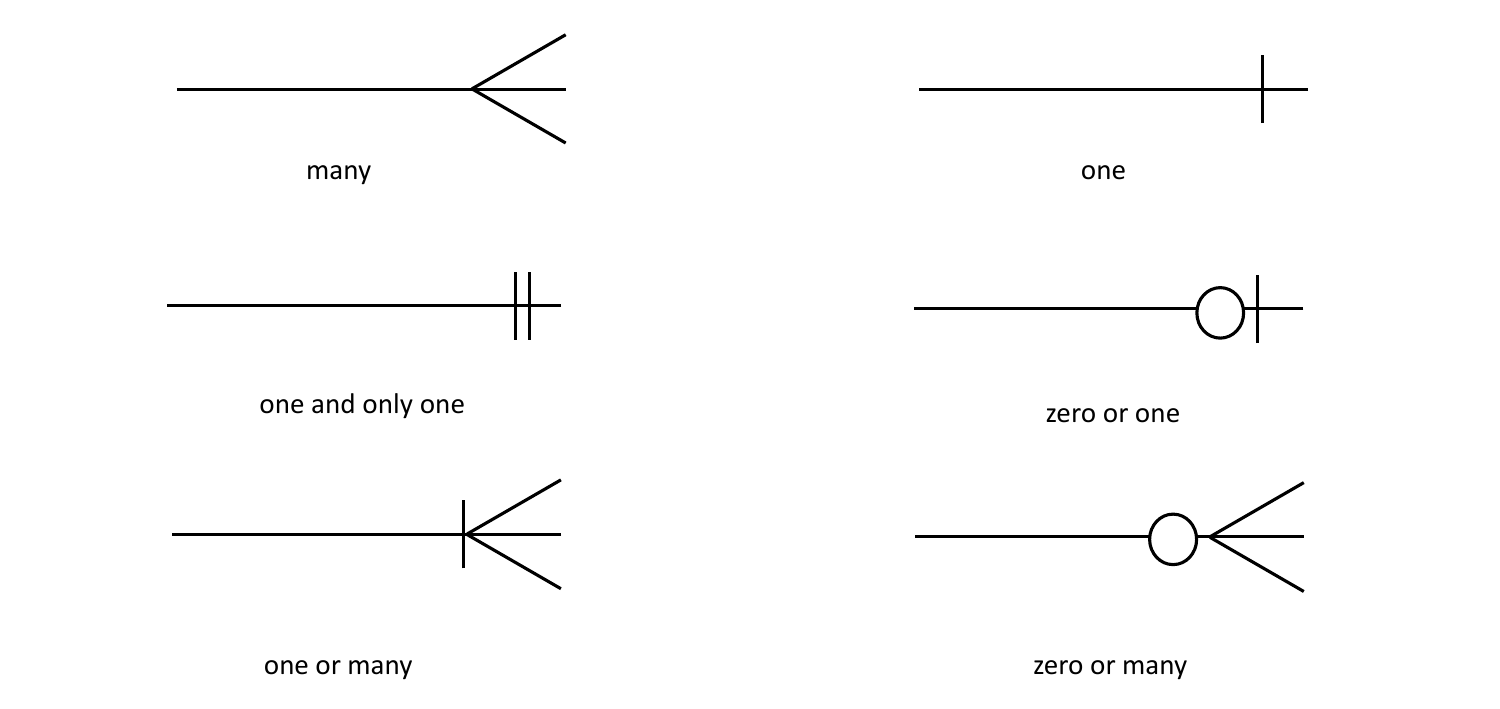}
	   \caption{Crow's foot notation}
	   \label{fig:crows-foot-notation}
    \end{figure}
    \clearpage
    \section{List of Mathematical Symbols Used}
    \label{appendix:mathematical-symbols}
    \begin{table}[!htb]
        \centering
        \begin{tabular}{|c|l|}
            \hline
             \textbf{Symbol} & \textbf{Description} \\
             \hline
             $i$ & Index of layer (Section~\ref{subsec:model-structure-params}) \\
             \hline
             \multicolumn{2}{|l|}{Security parameters} \\
             \hline
             $C_i$ & Criticality \\
             $S_i$ & Sensitivity \\
             \hline
             \multicolumn{2}{|l|}{Security maturity model} \\
             \hline
             $p_i$ & Practices \\
             $d_i$ & Domains \\
             $o_i$ & Objectives \\
             $m_i$ & Maturity \\
             $\mu_i$ & Maturity function \\
             \hline
             \multicolumn{2}{|l|}{Insurance pricing} \\
             \hline
             $\pi_i, v_i$ & Probability of loss\\
             $P_i$ & Premium \\
             $L_i$ & Loss value \\
             $u_i$ & Utility function \\
             \hline
             \multicolumn{2}{|l|}{Utility function parameters} \\
             \hline
             $\lambda_i$ & Loss estimation function \\
             $\Delta$ & Change in parameter \\
             $\alpha, \beta, \gamma$ & Insurer specific productivity parameters\\
             $\Lambda_i$ & Maximum potential loss per layer\\
             $\kappa_i$ & Policy limit\\
             $r_i$ & Premium rate \\
            \hline 
             
        \end{tabular}
        \vspace{1mm}
        \caption{List of mathematical symbols used in modelling framework}
        \label{tab:appendix-maths-symbols}
    \end{table}
\end{appendix}
\clearpage

\end{document}